\documentstyle[preprint,epsfig,aps]{revtex}
\begin{document}
\unitlength 1cm
\newcommand{\str}{\rule{0ex}{2.7ex}}
\newcommand{\strr}{\rule{0ex}{3.50ex}}
\newcommand{\strrr}{\rule{0ex}{4.50ex}}
\newcommand{\PP}{$A_{\pi\pi}$}
\newcommand{\PH}{$A_{\pi p}$}
\title{$\pi\pi$, $K\pi$ and $\pi N$ potential scattering and a
prediction of a narrow $\sigma$ meson resonance}
\author{ M.\ Sander  and H.V.\ von Geramb%
\footnote{e--mail: {\tt hvg@i04ktha.desy.de}}}
\address{Theoretische Kernphysik, Universit\"at Hamburg,
D--22761 Hamburg, Germany}
\date{\today}
\maketitle
\begin{abstract}
Low energy scattering and bound state properties of the
$\pi N$, $\pi\pi$ and $K\pi $ systems are studied as coupled
channel problems using inversion potentials of
phase shift data.
In a first step we apply the potential model to explain
recent measurements of pionic hydrogen shift and width. Secondly,
predictions of the model for pionium lifetime and shift 
confirm a well known and widely used effective range expression.
Thirdly, as extension  of this confirmation, we predict an unexpected
medium effect of the pionium lifetime 
which  shortens by several orders of magnitude.
The $\sigma$ meson shows a narrow resonance structure 
as a function of the medium modified mass with the implication of being
essentially energy independent. Similarly, we see this medium resonance
effect realized for the $K\pi$ system.
To support our findings we present also results 
for the $\rho$ meson and the $\Delta$(1232) resonance.
\end{abstract}

\pacs{12.39.Pn,13.75.Gx,13.75.Lb,14.40.Cs,36.10.Gv}

\section{Introduction}
It is difficult to name  hadronic systems for which the
underlying QCD structure is manifest in low and medium energy
observables \cite{barn97}. 
The reason for this are the significantly
different masses of nucleons and mesons as compared to the quark masses.
As a consequence we may
                    observe the QCD effects at very short interhadronic
distances with an energy scale for excitations
in the GeV  and not as it is customary for nuclear physics in the
MeV region. In the language of quantum mechanical
radial wave functions we anticipate
a situation similar to atomic physics where it is difficult to disclose
effects of the nuclear realm. 
The interplay between low energy nuclear physics and QCD is tough
since the phenomena are still difficult to identify and to associate with a
particular substructure.
Triggered by the need of a scalar isoscalar medium weight
$\sigma$ meson to understand the medium range nucleon--nucleon (NN)
attraction, there exist many attempts to localize a resonance
in $L=0,T=0$ $\pi\pi$ scattering with a mass $m_\sigma \sim 300-700$ MeV
\cite{har96,toe96}.
Despite all attempts,
experimental data do not support the anticipated narrow resonance
and theory gets acquainted with this situation.

We have been triggered by the development of a one solitary boson
exchange model (OSBEP) \cite{jae97} for NN scattering to study
the $\pi\pi$ and $\pi$--nucleon ($\pi N$) system in terms of a potential model.
The purpose of this investigation is to find a quantitative first 
guess for a dynamical model to  describe the resonant
particles $\rho,\;\sigma,\;\Delta$ etc. 
It is intended to treat them explicitly as interacting 
multi pion and nucleon systems
in an NN interaction model above meson production threshold.
Undoubtedly, such a model is a subtle many--body problem for supercomputing.
 
Despite this motivation,
there exists considerable interest in studies of di--hadronic systems
which form Coulomb bound states
\cite{sigg95,afa94}. Precision measurements of
shifts and widths of such hydrogen like two body systems 
are supposed to sense with high accuracy the strong interaction at low energy.
The shift is mainly caused by a modification at short distances where both the
Coulomb and hadronic interactions are present while 
the width results from decay into energetically
open channels. The decay can lead to a lower lying state of the same
system or into reaction channels.
Candidates for such studies are all oppositely charged
hadron pairs. 
From this sample, we selected pionic hydrogen (\PH) as a bound 
$\pi^-$--proton system and pionium (\PP) as a bound $\pi^-\pi^+$ system
both in relative s--states.

To describe such atomic systems it requires to know the hadronic interaction 
a priori or make a fit with a potential ansatz.
The hadronic interaction is known, in the sense that the partial wave 
phase shifts are sufficient
to determine the interaction by quantum inversion. This mathematical
method uses spectral theory 
to transform the boundary conditions in form
of the S--matrix or Jost function into a local energy independent potential
\cite{inv}. 
The radial Schr\"odinger and Klein--Gordon equations are 
equivalent to a Sturm--Liouville problem on the half axis
which is central for Gelfand--Levitan--Marchenko inversion. 
In Sec. II
we show the salient features of inversion and give in Sec. III 
the potential results for $\pi N$, $\pi\pi$ and $K \pi$ scattering.
With these potentials all ingredients are specified for the treatment 
of \PH\ and \PP\ as 
coupled channel problems and predictions are given in Sec. IV.
These predictions are in good agreement with experiment and lent support to our 
potential model.

As extension of the coupled channel potential model we treat $\pi N$, $\pi\pi$ 
and $K \pi$ scattering in the continuum. The coupled equations, with Coulomb 
effects included, confirm that isospin is hardly broken and that resonance 
properties of all systems are quantitatively well described. 

The p--wave resonances $ \Delta$ and $\rho$ are contained in Secs. III.B and
D respectively. 
S--wave channels of $\pi\pi$ and $K \pi$ scattering are treated in Sec. V. 
We attempt to solve the puzzle why a resonance is not visible 
in the phase shifts but at the same time meson exchange models require a 
medium weight $\sigma$ meson with definitive particle properties.
The solution to this investigation is a medium modification of the two body 
scattering which is manifest only in the $T=0$ $\pi\pi$ channel 
and $T=1/2$ $K \pi$ channel.
The medium modification is identified with an effective mass in the coupled
equations leading to a resonance structure
as a function of effective mass and not of energy. Thus we observe  a 
$\sigma$ resonance width $\Gamma \approx 1$ MeV practically independent of
energy. Associated with this $\sigma$ resonance is a shortening 
of lifetime of pionium \PP\ by more than three orders of magnitude. 
The conclusions in Sec. VI 
comprise the prediction of a very short range
attraction close to the origin and a narrow high repulsive barrier
at a relative distance $r \approx 0.2$ fm 
for the $\Delta$, $\rho$, $\sigma$ and 
$K_0^\ast$ resonances. Comparable short range repulsions 
are seen for other channels. A confirmation of the long range interaction 
in terms of meson exchange is given and most importantly a narrow
 $\sigma$ meson is predicted as a medium effect.

\section{Potentials from inversion}

Contrary to the direct path to obtain a potential for
elementary particle scattering from QCD or other
microscopic models we apply  quantum inversion to
experimentally determined  phase shift functions as input
in Gelfand--Levitan--Marchenko equations
\cite{inv}. Nowadays, the inversion techniques for nucleon--nucleon
scattering have evolved up to almost perfection for scattering
data below pion production threshold. Numerically, input phase shifts
can be reproduced for single and coupled channels
with a precision of $1/100$ of a degree, which is much
lower than the experimental uncertainty. This accuracy and the
possibility to test the inversion potential online,
i.\,e.\ inserting           the
potential into the scattering equation and reproducing the input phase
shifts, makes them a reliable and easy--to--handle tool for
quantitative medium energy nuclear physics. Guided by this spirit, the
utmost aim of quantum inversion is to provide a simple but
accurate operator to reproduce data. This paradigm, however, proscribes to
include sophisticated momentum dependencies or non--localities in the
potential since this requires more information then can be extracted
from phase shifts in a very limited energy domain.
Our studies and applications are limited  to the real
phase shift domain but, for mathematical reasons, they are
with smooth real functions extrapolated
towards higher energies and infinity. This extrapolation is  a
kind of regularization which does not introduce  spurious
low energy phenomena and thus has no effect upon results and
conclusions. This aspect of inversion has been investigated in
important circumstances and we assume to be on save ground
also in the applications studied hereafter. Furthermore, we are
facing only single channel situations without Coulomb effects.

The basic equation of inversion is the Sturm--Liouville equation \cite{levitan}
\begin{equation}
\left[ - {d^2 \over dx^2} + q(x) \right] y (x) = \lambda y(x).
\end{equation}
We use the equivalent radial Schr\"odinger equation
\begin{equation}
\left[ -{d^2 \over dr^2} + {\ell ( \ell+1) \over r^2 } +
{2 \mu c^2\over (\hbar c)^2} V_\ell (r)
\right] \psi_\ell (k,r) = k^2 \psi_\ell (k,r),
\qquad 0\le r < \infty
\label{rse}
\end{equation}
where $V_\ell (r)$ is a local, k--independent operator in
coordinate space and the factor $2\mu c^2/(\hbar c)^2$ guarantees the correct
units.
Boundary conditions for the physical solutions are
\begin{equation} \lim_{r\to 0} \psi_\ell(k,r)=0
\end{equation}
and
\begin{equation} \lim_{r \rightarrow \infty} \psi_\ell (k,r) =
\exp ( i {\delta_\ell (k)}) \sin (kr - {\ell \pi \over 2}
+ {\delta_\ell (k)})
\end{equation}

The Marchenko and the Gelfand--Levitan inversion
are two  inversion algorithm for the Sturm--Liouville
equation which are briefly sketched for single channels and
the case without a Coulomb reference potential.
More details can be found elsewhere \cite{inv,san97}.
\subsection{Marchenko Inversion}
The experimental information enters in the Marchenko inversion via the
$S$--matrix, which is related to the scattering phase shifts by the
relation
\begin{equation}
\label{sm}
S_{\ell}(k)=\exp(2i\delta_{\ell}(k)).
\end{equation}
We use a rational function interpolation and  extrapolation of
real data $\delta_\ell(k)$,  
\begin{equation} \label{delrat}
\delta_\ell (k) = \sum_{m=1}^M {D_m \over k - d_m}
\end{equation}
with the boundary conditions
\begin{equation}
\lim_{k\to0} \delta_\ell (k) \sim k^{2 \ell + 1} \qquad
\mbox{and} \qquad 
\lim_{k\to\infty} \delta_\ell (k) \sim k^{-1}.
\end{equation}
In any case there are 2--4 poles $d_m$ and strengths $D_m$  sufficient 
to provide a smooth description of data. 
Using a $[4/4]$ or $[6/6]$ Pad\'{e} approximation 
for the exponential function $e^z$ and
substituting the rational phase function (\ref{delrat}) 
into $z = 2i\delta_\ell(k)$ gives a rational S--matrix 
\begin{equation} \label{smrat}
S_\ell (k) = 1 + \sum_{n=1}^{2N} {s_n \over k - \sigma_n}
= \prod_{n=1}^{N} {k + \sigma_{n}^\uparrow \over k - \sigma_{n}^\uparrow}
\cdot {k + \sigma_{n}^\downarrow \over k - \sigma_{n}^\downarrow},
\end{equation}
using the notation
$
\{ \sigma_n^\uparrow \} := \{ \sigma_n | \mbox{Im} (\sigma_n) > 0 \}
$
and
$
\{ \sigma_n^\downarrow \} := \{ \sigma_n | \mbox{Im} (\sigma_n) < 0 \}.
$

The Marchenko input kernel 
\begin{equation} \label{minpk}
F_\ell (r,t) = -\frac{1}{2\pi} \int_{-\infty}^{+\infty} h^+_\ell(kr)
                              \left[ S_\ell(k)-1
 \right] h^+_\ell(kt) dk
\end{equation}
is readily computed with
the Riccati--Hankel functions $h^+_\ell(x)$ and contour integration.
This implies an algebraic equation for the translation kernel $A_\ell(r,t)$
of the Marchenko equation
\begin{equation} \label{mfeqn}
A_\ell (r,t)+F_\ell (r,t)+\int_{r}^{\infty}A_\ell(r,s)F_\ell(s,t)ds = 0.
\end{equation}
The potential is obtained from the translation kernel derivative
\begin{equation}
\label{potm}
V_{\ell}(r)=-2\frac{d}{dr}A_{\ell}(r,r).
\end{equation}
The rational representation of the scattering data leads to
an algebraic form of the potential \cite{inv}.

\subsection{Gelfand--Levitan Inversion}
Gelfand--Levitan inversion uses Jost functions as input. The latter is related
to the S--matrix by
\begin{equation}
\label{rhp}
S_{\ell}(k)=\frac{F_{\ell}(-k)}{F_{\ell}(k)}.
\end{equation}
Using the representation (\ref{smrat}), the Jost function in 
rational representation is given by
\begin{equation} \label{jost_1}
F_\ell (k) = \prod_{n=1}^{N} {k - \sigma_{n}^\downarrow \over
k + \sigma_{n}^\uparrow}
=
 1 + \sum_{n=1}^N {B_n \over k + \sigma_n^\uparrow},
\end{equation}
or
\begin{equation}
|F_\ell (k)|^{-2} = 1 + \sum_{n=1}^{N} 
{ L_n \over k^2 - \sigma_{n}^{\downarrow 2}}.
\end{equation}
The input kernel
\begin{equation}
G_\ell (r,t) = \frac{2}{\pi} \int_{0}^{\infty} j_\ell (kr) \left[
             \frac{1}{|F_\ell (k)|^2}- 1
 \right] j_\ell (kt) dk,
\end{equation}
$j_{\ell}(x)$ the Riccati--Bessel functions, is analytic.
The Gelfand--Levitan equation
\begin{equation}
K_\ell (r,t)+G_\ell (r,t)+\int_{0}^{r}K_\ell (r,s)G_\ell (s,t)ds = 0,
\end{equation}
relates input and translation kernels and the potential is defined by
\begin{equation}
\label{vgl}
V_{\ell}(r)=2\frac{d}{dr}K_{\ell}(r,r).
\end{equation}
For this potential an algebraic form is known \cite{inv}.

Gelfand--Levitan and Marchenko inversions 
yield the same potential. Numerical instabilities
can make the potentials differ but this signals in practice 
a problem and thus is permanently checked.
\section{Inversion potential results}

Today, partial wave phase shift analyses are available for many
hadronic systems. The best known of this sample is the
NN analysis of Arndt {\em et al.} (SAID) which covers an energy range 0--1.3 GeV
for $np$ and 0--1.6 GeV for $pp$ scattering \cite{arndt--said}. The $pp$ data
are presently extended up to 3 GeV with 
measurements from the EDDA Collaboration of COSY in J\"ulich 
\cite{Edda}. Of a similar
quality is the $\pi N$ analyses which is also available from
SAID. Partial wave phase shifts of the $\pi N$ system are determined
by an analysis of elastic  
$\pi^+ p \to \pi^+ p ,\;
\pi^- p \to  \pi^- p$ and charge exchange 
$\pi^- p \to \pi^0 n$ scattering. We use the solution SM95 of Arndt {\em et al.}
\cite{arnpin}. This and all other
analyses suppress Coulomb effects and
assume good isospin. This implies that mass differences between
$\pi^\pm$ and $\pi^0$  as well as proton and neutron are neglected.
As a rule, for the pion is used $m_\pi=m_{\pi^\pm}=139.5676$ MeV
and for the nucleon $m_N=m_p=938.27231$ MeV respectively.
Included is also the Karlsruhe--Helsinki analysis of Koch and
Pietarinen  KH80 \cite{koc80}.
For $\pi\pi$ scattering we use phase shifts from the analysis of
Frogatt  and Petersen \cite{fro77} and theoretical predictions
from a meson exchange potential by Lohse {\em et al.} \cite{loh90} and
chiral perturbation theory by Gasser and Leutwyler \cite{gas83}.
Finally, the $K\pi$ analysis of Estabrooks {\em et al.}
uses  final state interactions of
$K^\pm p \to K^\pm\pi^+ n$ and $K^\pm p \to K^\pm \pi^-  \Delta
^{++}$ \cite{esta78}. We restrict our analyses herein to $L=0$ and 1
partial waves with isospins $T=1/2$ and $3/2$ for
$\pi N$ and $K\pi$, and $T=0$,1 and 2 for the $\pi \pi$ systems.
A comprehensive analysis of data and inversion potentials
can be found elsewhere \cite{san97}.
We are using the elastic domain phase shifts and thus limit the input to
$T_{lab}< 500$ MeV for $\pi N$, $M_{\pi \pi} < 970$ MeV 
and  $M_{K\pi}< 1.3$ GeV. Resonance effects, like the $f_0(975)$
in $\pi\pi$ scattering, are not included.

\subsection{$\pi N$ s--wave scattering}

The notation of this channel distinguishes $S_{11}$ and $S_{31}$
partial waves to signal angular momentum $L=0$,
isospins $T=1/2$ and $3/2$ and spin $S=1/2$ states.
Input phase shifts from SM95 and KH80 are shown in Fig. 1 and
the inversion potentials are given in Fig. 
2 respectively.
In Sec. IV.A these potentials are used to study pionic hydrogen.

As a brief note we mention the assessment of the pion nucleon
coupling constant from these potentials.
From the $\pi N$ potentials in the
$T= 1/2$, $3/2$ s--channels we   find
the scattering lengths
$a_1   =  0.178 \, m_\pi^{-1}$ and $ a_3 = -0.088 \, m_\pi^{-1}$.
For a comparison with several other predictions
see Table I.
These results may be used in the Goldberger--Miyazawa--Oehme
sum rule \cite{wor92}
\begin{equation}
{\displaystyle f_{\pi NN}^2 \over \displaystyle 4 \pi} =
{\displaystyle  (m_\pi^2 - \mu^2) ( m_N + m_\pi) \over
\displaystyle 6 m_N  m_\pi}
( a_1 - a_3) -
{\displaystyle  m_\pi^2 - \mu^2 \over \displaystyle 8 \pi^2}
\int_0^\infty {\displaystyle \sigma_{\pi^-p} - \sigma_{\pi^+p}
\over \sqrt{ \displaystyle q^2 + m_\pi^2}} dq
\end{equation}
to obtain a model independent estimate for the $\pi NN$ coupling constant.
Using the simplified form of this sum rule 
\begin{equation}
{\displaystyle f_{\pi NN}^2 \over \displaystyle 4 \pi} =
 0.19 m_\pi ( a_1 - a_3) - (0.025 \mbox{ mb}^{-1})
{\cal J},
\end{equation}
where the integral
\begin{equation}
{\cal J} = {\displaystyle  1 \over \displaystyle 4 \pi^2}
\int_0^\infty {\displaystyle \sigma_{\pi^-p} - \sigma_{\pi^+p}
\over \sqrt{\displaystyle q^2 + m_\pi^2}} dq
\end{equation}
has the VPI value
$
{\cal J} = -1.041 \mbox{ mb}.
$
We find
\begin{equation}
{\displaystyle f_{\pi NN}^2 \over \displaystyle 4 \pi} = 0.0766,
\qquad \mbox{or} \qquad
{\displaystyle g_{\pi NN}^2 \over \displaystyle 4 \pi} = 13.84,
\end{equation}
which is fully consistent with the value of $13.75 \pm 0.15$ given in
\cite{arnpin}.

\subsection{$\pi N$ p--wave scattering}

Most prominent is the $\Delta(1232)$ or
$P_{33}$ resonance which we treat with great care.
The low energy phase shift function, shown in Fig. 3, uses 
SM95 and KH80 data.
A factorization of the S--matrix into
a resonant and a non--resonant background part
$S (k) = S_r (k) S_b (k)$ is useful.
For the resonant part 
a resonance and an auxiliary pole parameterization
\begin{equation} \label{sresonance}
S_r (k) =
{ ( k + k_r ) ( k - k_r^* ) \over (k - k_r ) ( k + k_r^* ) }
{ ( k + k_h ) ( k - k_h^* ) \over (k - k_h ) ( k + k_h^* ) }
\end{equation}
is used. It contains the right amount of zeros and poles for a
decomposition into Jost functions.
The background  $S_b (k)$  is smooth but its shape depends on the
parameters in (\ref{sresonance}).
Actual values are taken from data tables \cite{PDG96} for $k_r$ and
$k_h=0.5+i 10$ fm$^{-1}$.
The inversion
potential is independent from this splitting and is shown in Fig. 4.
The resonance feature is generated from the short range 
attraction near the origin and is limited by
a narrow (0.1 fm) potential barrier with a height of
$\sim 20$ GeV. 
Being accustomed to strengths and ranges of 
NN potentials it is surprising to see this small radial
dimension and large potential strength. 
After a second thought this is not a surprise since inversion is a kind of
generalized Fourier transformation and thus the units are a consequence of the
pion mass and the quantitative behavior of the phase shift.
Nevertheless, the potential changes occur at very small radii 
compared to the size of the charge form factors of pion and nucleon.
The RMS radii are known to be 0.54 fm for the pion and
0.7 fm for the nucleon.
Our radius describes the distance of the center of masses and $r=0.16$ fm,
the barrier radius,  
implying more than 90\% overlap of the intrinsic structures.
We conjecture for the barrier a simulation of a transition from the
pion--nucleon quark content into the 3--quark content of the $\Delta$.
Ultimately, such explanation must be confirmed by QCD calculations.

In Fig. 5 we illustrate the $P_{33}$ potential resonance
showing the physical
solution $|u(E,r)/r|^2$ as a function of energy and radius.
This figure shows that the probability builds up between the origin 
and the barrier. 

The long range part of the inversion potential,
not visible in Fig. 4, behaves like a Yukawa tail with a strength
$Y=650.0$  MeVfm and  an effectively exchanged mass of 350 MeV.
There exists no physical particle with this mass since in the
meson exchange picture s-- and t--channel graphs contribute.

\subsection{$\pi\pi$  s--wave scattering}

$\pi\pi$ phase shifts come from the analysis of final
state interactions in $\pi N \rightarrow \pi \pi N $
systems or the $K_{e4}$--decay
$K^- \rightarrow \pi^+ \pi^- e \bar{\nu}$.
Here we use results of the CERN--Munich experiment \cite{fro77}.
The scattering is purely elastic until $M_{\pi\pi} = 987.3$ MeV where
coupling to the $K \bar{K}$ opens and the phase shift becomes
highly inelastic and resonant in the $T=0$ channel. The $T=2$
channel remains smooth. A summary of all experimental phase shifts
and theoretical predictions, from a meson exchange  model
\cite{loh90} and chiral perturbation theory \cite{gas83}, is shown
in Fig. 6. Notation for the isospin and angular momentum channels uses
$\delta_\ell^T \mbox{or } V_\ell^T$.The three used phase shift
sources yield slightly different
inversion potentials, shown in Fig. 7. 
It is obvious
that $\pi\pi$ phase shifts are less well established 
than $\pi N$ data. Table II
comprises  a summary of effective range parameters from different
sources to substantiate the uncertainties.

\subsection{$\pi\pi$  p--wave scattering}

As before, phase shifts come from the analysis of final
state interactions in $\pi N \rightarrow \pi \pi N $
systems. Isospin is limited to
the single value $T=1$ and the
CERN--Munich  analysis is used \cite{fro77}.
The scattering is dominated by the $\rho$ resonance, $m_\rho=770$ MeV,
$\Gamma=150$ MeV,  and the phase shift remains essentially real
until $M_{\pi\pi} = 1.2$ GeV. The used phase shift is shown
in Fig. 8 and the inversion potential is shown in Fig. 9. 
Notice the barrier maximum at $0.16$ fm which is the same as seen in 
the $\pi N$ $P_{33}$ channel.
We decline from repeating the resonance wave function display due to
its similarity with results shown in Fig. 5.

\subsection{$K\pi$ s--wave scattering}

Phase shifts are taken from the Estabrooks {\em et al.}
analysis \cite{esta78}. We distinguish between isospins
$T=1/2$ and $3/2$. As shown in Figs. 10 and 11,
this systems resembles the $\pi\pi$ s--wave scattering. The long range part of
the two isospin potentials are numerically very close outside
$r=0.4$ fm  which supports isospin independence. Furthermore,
outside  $r=0.8$ fm both potentials are very small,  $|V_0^{2T}(r)|
< 1$ MeV. This weak medium and long range interaction requires
further investigations but we  notice the same feature for
$K^+ N$ inversion potentials,
consistent with the extraordinary long
mean free path of kaons in nuclear matter \cite{dover82,san97}.
We shall show that this system displays 
similar medium effects as the corresponding $\pi\pi$ system.

\section{Coupled systems}

Pionic-hydrogen \PH\ is treated  in a coupled channel calculation as a
charge exchange resonance  seen in  $\pi^0$--neutron  scattering.
There exists a spectrum of excited states with different relative
angular momenta but only the s--states shall be studied for comparison
with experiment. Similarly, 
\PP\ is formed by $\pi^-\pi^+$ and it is treated as a charge
exchange resonance state seen in the elastic $\pi^0\pi^0$  channel.

The coupled system is written in the form
\begin{equation} \label{eqn1}
f_i^{\prime\prime} + k_i^2 f_i = \sum_{j=1}^2 {2\mu_i\over \hbar^2} V_{ij} f_j,
\end{equation}
and uses conventionally $i=1(2)$ as the reaction (elastic entrance)
channel. 

\subsection{Pionic hydrogen}
The reduced masses enter in our calculation with two options. In the first case 
we use only the mass of the charged pion and the proton mass, $\mu_1 = \mu_2$ 
with
\begin{equation}
\mu_1={m_{\pi^-}m_p\over (m_{\pi^-}+m_p) }.
\end{equation}
In the second case we use $\mu_1 \neq \mu_2$ and use for $\mu_2$ the 
physical masses of the neutral particles
\begin{equation}
\mu_2={m_{\pi^0}m_n\over (m_{\pi^0}+m_n) }.
\end{equation}
The potential matrix contains the Clebsch--Gordan coefficients for the rotation 
of the interaction from good isospin into particle states.
\begin{eqnarray}
 V_{11}  & = &  V^{\pi^-p} =
\frac13 V_s^{3/2} + \frac23 V_s^{1/2} - \frac{e^2}{r}, \\
V_{12}  = V_{21}  & = & {\sqrt{2}\over 3}(V_s^{3/2} - V_s^{1/2}), \\
V_{22}  & = & V^{\pi^0 n} = \frac23 V_s^{3/2} + \frac13 V_s^{1/2}.
\end{eqnarray}
The kinematics  is expressed in any case by the 
physical masses and projectile kinetic energy $T_{lab}$
\begin{eqnarray}
 S&=&(m_{\pi^0}+m_n)^2 +2T_{lab} m_n  \\
 k^2_1&=&{S^2+(m_{\pi^-}^2-m_p^2)^2-2S(m_{\pi^-}^2+m_p^2)
       \over 4 S(\hbar c)^2 } \\
 k_2^2&=&{m_{n}^2T_{lab}(T_{lab}+2 m_{\pi^0})\over
               S  (\hbar c)^2 },
\end{eqnarray}
which guarantees the correct threshold behavior.

The Coulomb attraction between $\pi^-$ and $p$ causes a
bound system  which is known as
pionic hydrogen \PH. In addition to Coulomb attraction,
the hadronic interaction between
the two constituents distorts the short range interaction and
changes the pure Coulomb spectrum.
Decay channels are
$ \pi^-  p \to \pi^0 n + 3.30023$ MeV and $ \pi^- p \to n\gamma $.
The hadronic shift of the $3p \to 1s$ transition and the total
$1s$ width has been measured
at PSI \cite{sigg95}. To analyze this experiment we
use the described potential model.
Inelasticities, Coulomb and other isospin breaking effects are
supposed to be not included in the SM95 phase shift analysis
and the real potential matrix is shown in Fig. 12.
We use the reduced masses $\mu_1=\mu_2 = 121.4970$ MeV
which are consistent with the phase shift analysis and using
$\pi^\pm$ and $p$ masses. Alternatively $\mu_1=121.4970$ MeV and
$\mu_2=118.0216$ MeV based upon $\pi^0$ and neutron masses respectively
can be used.

The bound states of the  $\pi^-p$ system can be found as resonances
in the
energetically open $\pi^0n$ channel. The width (FWHM) of this resonance
accounts only for the decay of
$\pi^-p\to\pi^0n$. Table III contains this result in the first line.
To evaluate the hadronic shift we used in all cases the reference
energy $E_{1s}^C=3234.9408$ eV. The shift is calculated as the
difference between $E_{1s}^C$ and the calculated resonance energy
taken at the maximum of the elastic scattering cross section
\begin{equation}
\sigma(\pi^0n \to \pi^0n)={\pi\over k_2^2}|1-S_{22}|^2.
\end{equation}
From this distribution we obtain also the FWHM. The Coulomb potential
is contained in $V_{11}$ as determined from point charges $V^C=e^2/r$
or double folded Gaussian charge distributions $V^C=e^2 \Phi(1.13r)/r$.
$\Phi(r/\alpha )$ is the error function and $\alpha=\sqrt{<r^2_\pi>+
<r^2_p>}$ with RMS radii $<r^2_\pi>^{1/2}=0.5389$ fm and $<r^2_p>^{1/2}
=0.702$ fm.
To compare with experimental data \cite{sigg95} we evaluated the
partial width $\Gamma^{\pi^0n}_{1s}$ using the
Panofsky ratio $P =1.546 \pm 0.009$ \cite{spu77},
\begin{equation}
\Gamma = \left(
1 + \frac1P \right) \Gamma^{\pi^-p\to\pi^0n}.
\end{equation}
To account also for decay into
the $n\gamma$ channel from the $\pi^0n$ channel an imaginary potential
$W_{11}=-9\exp(-9r^2)$ MeV is added to $V_{11}$ which brings the
theoretical results in agreement with experiment. Other corrections are
not further pursued since they enter into the reference and
resonance energy
with approximately the same amount, affecting the difference by
$<\pm 0.01$ eV. We shall not dwell upon this issue in more detail
herein.
The essential result of this part of our calculation is that
isospin breaking effects due to mass differences are small.

\subsection{Pionium}
Very close the same convention as for \PH\
is used here. Only one value of a reduced 
mass occurs 
\begin{equation} \label{refmass}
\mu_1=\mu_2={m_{\pi^+}\over 2}.
\end{equation}
This value is varied in Sec. V.A when medium effects are discussed. The 
potential matrix is readily expressed by
\begin{eqnarray}
 V_{11}  & = &  V^{\pi^+\pi^-} =
\frac13 V_0^2 + \frac23 V_0^0 - \frac{e^2}{r}, \\
V_{12}  = V_{21}  & = & {\sqrt{2}\over 3}(V_0^2 - V_0^0), \\
V_{22}  & = & V^{\pi^0\pi^0} = \frac23 V_0^2 + \frac13 V_0^0,
\end{eqnarray}
and the kinematics by
\begin{eqnarray}
 S&=&4m_{\pi^0}^2 +2T_{lab} m_{\pi^0} \\
 k^2_1&=&{S-4m_{\pi^-}^2
       \over 4 (\hbar c)^2 } \\
 k^2_2&=&{m_{\pi^0}^2T_{lab}(T_{lab}+2 m_{\pi^0})\over
               S  (\hbar c)^2    }  \\
 M_{\pi\pi}&=&\sqrt{S}
\end{eqnarray}

Similar to the \PH\ system there exists pionium \PP\
which is formed by $\pi^- \pi^+$ Coulomb attraction. It decays
predominantly by charge exchange into the open $\pi^0 \pi^0$ channel.
The coupled channel system is defined with Eqn. (\ref{eqn1}).
We assume the same approach as for  \PH\
and rotate the good isospin potentials into
particle states. The hadronic potential matrix is shown in Fig. 13
using the three different sources discussed in Sect. III.C.
Phase shift analyses and inversion use a single mass
$\mu_1=\mu_2 =\mu= m_{\pi^+}/2$
without Coulomb effects. This assumption guarantees good
isospins $T=0$ and 2. The results of our calculations are summarized
in Table IV. A point Coulomb reference energy $E_{1s}^C=
1.85807248$ keV  is used.  With this choice of masses and the correct
Q-value our calculations agree practically with some well known
scattering length expressions which is most often used \cite{afa94,efimov,ure61}
\begin{eqnarray}
{1 \over \tau} & = &
{8 \pi \over 9} \left( {2 \Delta m \over \mu} \right)^{\frac12}
{ \left(a_0^0 - a_0^2\right)^2 \, | \Psi (0)|^2
\over 1 + \frac29 \mu\Delta m\left(a_0^0 + 2 a_0^2\right)^2 }
\\
& \approx & 1.43 \left(a_0^0 - a_0^2\right)^2 \, | \Psi (0)|^2,
\end{eqnarray}
where $\Delta m$ is the mass difference $m_{\pi^+} - m_{\pi^0}$.
The real interesting result of this study comes from dramatic changes
(shortening) of the \PP\ lifetime, by several orders of magnitude,
with small variations of the reduced masses $\mu_i$.

\section{Medium effects}

Low energy nuclear physics associates medium effects with changes of
the free
    interaction potential or free two body scattering amplitudes
in the presence of a few-- or many--body environment. Effective
interactions incorporate such effects and degrades the environment
into the role of spectators. A well known example is the nuclear matter g--matrix
which includes Pauli blocking and selfconsistent mean field effects
and which is related to the free two body t-matrix. Other medium
modification are due to recoil,  truncation
of coupled channels  and relativistic corrections to name a few.
Medium effects from meson  and boson exchange models are related to
restoration mechanism of the broken chiral symmetry in nuclear matter.
This field of research is presently in the center of different
theoretical lines of thought and some of them are found in
\cite{rap96}.

Our concern is an effective mass in the two particle wave equation
which may have different causes  in few and many body systems. At
this stage the consequences and dynamical effects are shown
for the two particle subsystems $\pi\pi$ and $K\pi$
in case an effective mass is assumed.
What we study are partial wave phase shifts  $\delta(E,m)$ as a function of
energy and the effective mass, which determines the
reduced masses $\mu_i$ in Eqn. (\ref{refmass}). Effects upon the lifetime
of pionium are also studied.

\subsection{$\pi\pi$ scattering}

In Sect. IV.B results of pionium lifetime and hadronic shift are given.
This study uses the $\pi^\pm$ mass consistently with the phase shift
analyses and the scattering length expressions for the life time.
Here we extend this study and show results in Fig. 14
for the hadronic shift and lifetime  of pionium as a function
of the pion mass, $m_\pi=2\mu_1=2\mu_2$ in Eqs. (\ref{refmass}). The potential
matrix $V_{ij}$ was unchanged
and the $k_i^2$  values computed with the physical masses of $\pi^\pm$
and $\pi^0$ as defined in  Eqs. (38--41). With this prescription we
effectively change the total strength of the potential matrix $V_{ij}$
by $\pm1\%$ and observe a variation of the shift by two orders of
magnitude and of the lifetime by three orders of magnitude. The
physical mass result is accentuated. The interval of mass
variation is small compared with the physical mass
differences between $\pi^\pm$ and $\pi^0$ of 4.6 MeV, which a
correct phase shift analysis should consider. Since this is not the
case, consistency requires to use $\mu_1=\mu_2$ in the inversion
procedure and in the coupled channel calculations. Thus, we maintain
the assumption of a good isospin also for the calculation with
an effective mass.

Triggered from this result we extended our calculations for the
eigenchannel phase shifts of the coupled system  which coincide
with the uncoupled good isospin calculations for $T=0$ and 2.
In Fig. 15 we show the $L=0,\; T=0$ phase shift $\delta^0_0(m,T)$
as a function
of the effective mass $m=2\mu_1=2\mu_2$ and the laboratory kinetic
energy. The physical mass result is emphasized and follows
the result shown in Fig. 6 for $\delta^0_0$. As a function of mass
we observe a typical resonance behavior whose width $\Gamma=1-2$ MeV
which is almost energy independent. The resonance feature is
supported by Fig. 16 which shows $|u(r,m,T)/r|^2$ as a function
of radius and  effective mass for three kinetic energies. Since
$L=0$, the radial wave function is $\neq 0$ at the origin around
which the probability is build up. This figure should be compared
with Fig. 5 of the $\pi N$ $\Delta$ resonance which shows the
same pattern as a function of energy. For the $\pi N$ $P_{33}$
phase shift $\delta(E)$ varies only insignificantly when the reduced
mass of pion--nucleon system is modified within  1 MeV and
the resonance pole is stable against this variation. From the
$\pi N$ potential in Fig. 4 the same mass dependence as for the
$\pi\pi$ system was expected. This is not the case. We attribute this
stability to the boundary condition of the p--wave at the origin
to be zero. The $\pi\pi$ s-wave $T=0$ potential in Fig. 7 supports
the resonance as function of mass since no  boundary
condition at the origin restricts the development of  a large
amplitude breathing mode. The potentials have in both cases a very
high barrier of  20 -- 30 GeV, compared to the kinetic
energy of several 100 MeV, which is narrow $\sim 0.1$ fm and
within the barrier the potentials are very deep. For the s-- and
p--waves decays the radial wave exponentially practically
independent from the projectile energy. However,
the boundary condition of the p--wave at the origin suppresses the
free unfolding of a resonance amplitude  independent from energy.
It requires an optimal matching of the external wave function
to realize the internal resonance enhancement. A small variation
of the potential depth within the barrier does not overcome the
restriction from the boundary condition for the p--wave. This situation
is essentially different for s--waves. For them it is important to
have the correct wave number within the barrier that the wave function
matches  the
exponentially decaying function in the barrier optimally.
The wave number within the barrier is determined by
\begin{equation}
k^2={(M_{\pi\pi}-(2\mu/ m_{\pi^+})V(r))^2-4m_{\pi^+}^2
   \over 4(\hbar c)^2}.
\end{equation}
A small variation of the reduced mass $\mu$ produces the discussed
effect. In connection with the very high barrier gives this the
explanation of the projectile independence of the observed
$\pi\pi$ resonance as a function of the effective mass. The
high barrier decouples for s--waves the inner from the exterior
dynamics which  is not the case for p-- and higher partial waves.

Finally, the effective mass used in this study must be the result
of embedding the $\pi\pi$ system into a few-- and many--body
environment. In the meson exchange models for NN scattering should
the correlated two pion exchange dynamics not be determined from
the physical free pion mass but from an effective mass. This
mechanism is able to change the non--resonant two pion system into
a resonant system with a width $1 <\Gamma< 600$ MeV. The lower limit
is extracted from Fig. 15 and the upper limit is deduced from the
free system $\delta^0_0(e)$ \cite{har96}. This gives a possible
explanation of the properties of the $\sigma$--meson used in all
high quality meson echange model for hadron--hadron interactions
out of which the NN potentials are the best examples.
The phase shift $\delta^2_0(E,m)$ shows practically
no mass dependence and maintains its repulsive nature.

To support this picture for $\pi\pi$ scattering it is obvious to
look for other systems with the same properties.

\subsection{$K\pi$ scattering}

Phase shift data for the $K\pi$ s--wave scattering yield
qualitatively the same situation with two possible isospin couplings
$T=1/2$ and $3/2$. The inversion potentials are shown in Fig. 11.
A purely repulsive potential is seen for the $T=3/2$ channel whereas
$T=1/2$ has a narrow and high potential barrier at about the same
radial region as the $\pi\pi$ system and we expect great similarities
with respect to the phase shifts $\delta_0^1(m,T)$, a functions of
effective mass and kinetic energy. Fig. 17 confirms this conjecture.
Following the same line for the coupled system in Sec. IV, one 
effective mass parameter is used $m=2\mu_{K\pi}$ where the physical
mass  has a value of 215.94 MeV. The solid line signals the physical
$T=1/2$ phase shift and the energy--mass distribution shows
a resonance structure. The width increases, $\Gamma>80$ MeV, with
laboratory kinetic energy which is caused by the relative smallness
of the barrier of  800 MeV. The phase shifts $\delta^3_0(E,m)$
shows no mass dependence and maintains its repulsive nature.

\section{Summary}
The elastic scattering domain for $\pi\pi$, $K\pi$ and $\pi N$ scattering 
is investigated with the help of a local r--space potential model.
Quantum inversion is used to generate from phase shift data the potentials 
for low partial waves and the permitted isospin channels. 
Tables of these potentials are available \cite{www}. All resonance
features visible in the phase shifts are well accounted for, 
$\rho$ and $\Delta$ are potential resonances of a $\pi\pi$ and $\pi N$
system respectively.

Pionic hydrogen \PH\ is studied as a resonance in elastic $\pi^0$--neutron 
scattering below $\pi^-$--proton threshold. 
The calculated shift and width of the ground state agree very well 
with recent measurements. Similarly, pionium \PP\ is studied as a resonance 
in elastic $\pi^0\pi^0$ scattering below $\pi^+\pi^-$ threshold. 
The potential model confirms well known scattering length expressions 
often used with hadronic bound state problems. 
As a new feature we disclose a medium effect 
of lifetime and shift of \PP\ when the interaction strength is increased by
typically half a percent. The amount of change is more than three orders of
magnitude.
In an extended study a medium resonance is identified with the $\sigma$ 
resonance. In particular it is shown that this resonance is a function of the
effective mass, which we identify as the factor in front of the potential,
and not as it is common as a function of energy. 
It can be considered as a parametric resonance. The change of \PP\ lifetime 
is a feature of this resonance. A similar situation with medium
effects and a parametric resonance is identified in the 
$K \pi$ $L=0,\;T=1/2$ channel. Contrary to this medium resonances show the
classic $\rho$ and $\Delta$ resonances practically no effective mass dependence.
Showing radial wave functions we explain these differences of dynamical 
behavior with boundary conditions at the origin. All resonances are associated
with a small radial region within 0.3 fm. 
This distance is the center of mass distance of the two particles and  
is not the QCD bag radius. Table V shows that
the potential barriers which trap the relative system are very close 
the same with the exception of the strange $K \pi$ resonance.  
In view of the large RMS radii of mesons and nucleons 
($\sim 0.6$ fm) this implies
a more than 90 percent overlap of the individual QCD bags before  fusion 
occurs and the new configuration develops. 
We see in this universal small barrier radius
 a reason why meson exchange works and permits quantitatively 
to describe NN and other hadron--hadron interactions.
This finding requires detailed calculations on the QCD level.

For low and medium energy nuclear physics this potential model defines 
essential ingredients of boson exchange models in terms of $\pi$ mesons
and nucleons only. $\rho$ and $\Delta$ resonances are well accounted for 
their properties with the inversion potentials and the $\sigma$ meson
shows dynamic aspects which does not support it as a particle.  
Since long there exist experimental data which involve two pion production
and still lack of satisfactory explanation. It is known as ABC effect
\cite{abc61} and we suggest a re--analysis of this data with the discussed
dynamical effects included. 
The set of inversion potentials can easily be completed \cite{san97} to
have a basis for potential model calculations which use only pions, protons 
and neutrons. This opens the possibility for quantitative fragmentation 
calculations requiring massive supercomputing. 

\section*{Acknowledgment}
The authors appreciate the constructive discussions with R. Jahn, 
H. Markum, D. Bugg, L. J\"ade and G. Steinbach.
This work was supported in part by Forschungszentrum J\"ulich GmbH
under Grant--Nr. 41126865.


\begin{table}\centering
\caption[$\pi N$ s--Wellen Streul\"angen]%
{$\pi N$ s--wave scattering lengths and $\pi NN$ coupling constant
obtained from the GMO sum rule.}
\label{tab_one}
\begin{tabular}[htb]{cccccccccc}
& Model & \hspace{0.4cm} & $a_1$ [$m_\pi^{-1}$] & \hspace{0.4cm} &$a_3$
[$m_\pi^{-1}$]&\hspace{0.4cm}&$f_{\pi NN}^2/4\pi$&\hspace{0.4cm}&
  Ref. \\[0.1cm]
\hline
\str &  SM95 Inversion & & 0.178  & & $-0.088$ & & 0.0766& &  \\
\str & KH80 & & 0.173 & & $-0.101$ & & 0.079 & & \protect\cite{koc80} \\
& $\pi^-p$ 1s state & &  0.185 & &  $-0.104 $ & & 0.081 & &
\protect\cite{sigg95} \\
\str & Pearce et al.& & 0.151 & & $-0.092$ & & 0.072 & &
\protect\cite{pea91} \\
 & Sch\"utz et al. & & 0.169 & & $-0.085$  & &  0.074 & &
\protect\cite{sch94a} \\
\end{tabular}
\end{table}

\begin{table}\centering
\caption%
{$\pi\pi$ s--wave scattering lengths.}
\label{tab_two}
\begin{tabular}[thb]{ccccc}
 Source/Model & $a_0^0 \quad [m_\pi^{-1}] $ & $a_0^2 \quad
[m_\pi^{-1}] $ & $(a_0^0 - a_0^2) \quad [m_\pi^{-1}] $&Ref. \\[0.1cm]
\hline
\multicolumn{5}{c}{\em \strr Predictions from theory} \\
 Weinberg & 0.16 & $-0.046$ & 0.206 & \protect\cite{wei66} \\
$\chi$PT & 0.20 & $-0.042$ & 0.242 & \protect\cite{gas83} \\
Meson exchange & 0.31 & $-0.027$ & 0.337 & \protect\cite{loh90} \\
\multicolumn{5}{c}{\em \strr Results from experiment} \\
 $K_{e4}$ & $0.26 \pm 0.05$ & $-0.028\pm0.012 $ & $0.288 \pm 0.051$ &
\protect\cite{ros77} \\
Chew--Low PSA & $0.24 \pm 0.03$ & $-0.04 \pm 0.04$ & $0.280 \pm 0.050$&
\protect\cite{ale82} \\
Soft--Pion &  $0.188 \pm 0.016$ & $-0.037 \pm 0.006$ & $0.225 \pm 0.017$ &
\protect\cite{bur91} \\
$\pi N \to \pi\pi N$ & $0.205 \pm 0.025$ & $ -0.031 \pm 0.007$ & $0.236 \pm 
0.026$ & \protect\cite{ols97} \\ 
\multicolumn{5}{c}{\em \strr Results from inversion} \\
Froggatt phase shifts & 0.31 &  $-0.059$ & 0.369  & \\
$\chi$PT phase shifts & 0.20 &  $-0.043$ & 0.243  & \\
Meson ex. phase shifts & 0.30    &  $-0.025$ & 0.325  & \\
\end{tabular}
\end{table}

\begin{table}\centering
\caption{\PH\  $1s$ level shift with respect to $E_{1s}^C= 3234.9408$ eV and
width. The strength of the imaginary $W_{11}$ was adjusted to
reproduce the experimental value.}
\label{tab_pinb1s}
\begin{tabular}[b]{cccccccc}
 \str & \hspace{0.6cm} &\multicolumn{2}{c}{Point Charge Coulomb}&
\hspace{0.6cm} & \multicolumn{2}{c}{Gaussian Charge Coulomb}&
\\   
Mass & & Shift [eV] & FWHM [eV] & & Shift [eV] & FWHM [eV] &  \\
\hline
$\mu_1=\mu_2$  & & --7.29821 & 0.5250 & &  --7.16746 & 0.5230  &    \\
$\mu_1\neq\mu_2$ & & --7.13259 & 0.5187 & & --7.01055 & 0.5144 &\\
\multicolumn{7}{c}{ Shift =
--7.127$\pm$0.046  [eV],
$\Gamma_{1s}^{\pi^0n}$ = 0.590 [eV] }
& Sigg \cite{sigg95}
\\
$\mu_1\neq\mu_2$
& & --7.23259 & 0.9763  & & --7.12387 & 0.9763 & $V_{11}+iW_{11}$
\\
\multicolumn{7}{c}{ Shift =
--7.127$\pm$0.028$\pm$0.036  [eV],
$\Gamma_{1s}$ = 0.97$\pm$0.10$\pm$0.05 [eV]}
& Sigg \cite{sigg95} \\
\end{tabular}
\end{table}

\begin{table}\centering
\caption{\PP\ properties from inversion potentials}
\label{table2}
\begin{tabular}[b]{ccccc} $E_{1s}$ [keV] &
\multicolumn{1}{c}{Shift  [eV]}  &
\multicolumn{1}{c}{$\tau$ [$ 10^{-15}$ s]} & FWHM  [eV] &
 Ref. \\
\hline
   1.8638814 &-5.809  & 1.97   & 0.3481 &  Froggatt \protect\cite{fro77}   \\
   1.8635114 &-5.439 &  1.89  & 0.3627 &  Lohse  \protect\cite{loh90} \\
   1.8616174  &-3.545 & 3.22  &0.2128 &  $\chi_{PT}$  \protect\cite{gas83} \\
\multicolumn{5}{c}{\em  Predictions from experimental analysis
and other models} \\
   1.858 &   &  2.9$^{+\infty}_{-2.1}$ &  & Afanasyev
\protect\cite{afa94} \\
 1.865 & -7.0    &  3.2 & & Efimov  \protect\cite{efimov}
\end{tabular}
\end{table}

\begin{table}\centering
\caption{Potential barrier positions.}
\label{tab_five}
\begin{tabular}[htb]{ccccc}
Channel & $\pi N$--$\Delta$ & $\pi\pi$--$\sigma$ & 
$\pi\pi$--$\rho$ & $K\pi$--$K_0^\ast$ \\
Barrier radius [fm] &    0.165 &  0.145 & 0.152 & 0.303 \\
\end{tabular}
\end{table}


\begin{figure}\centering
\begin{picture}(15.0,15.0)(0.0,0.0)
\epsfig{figure=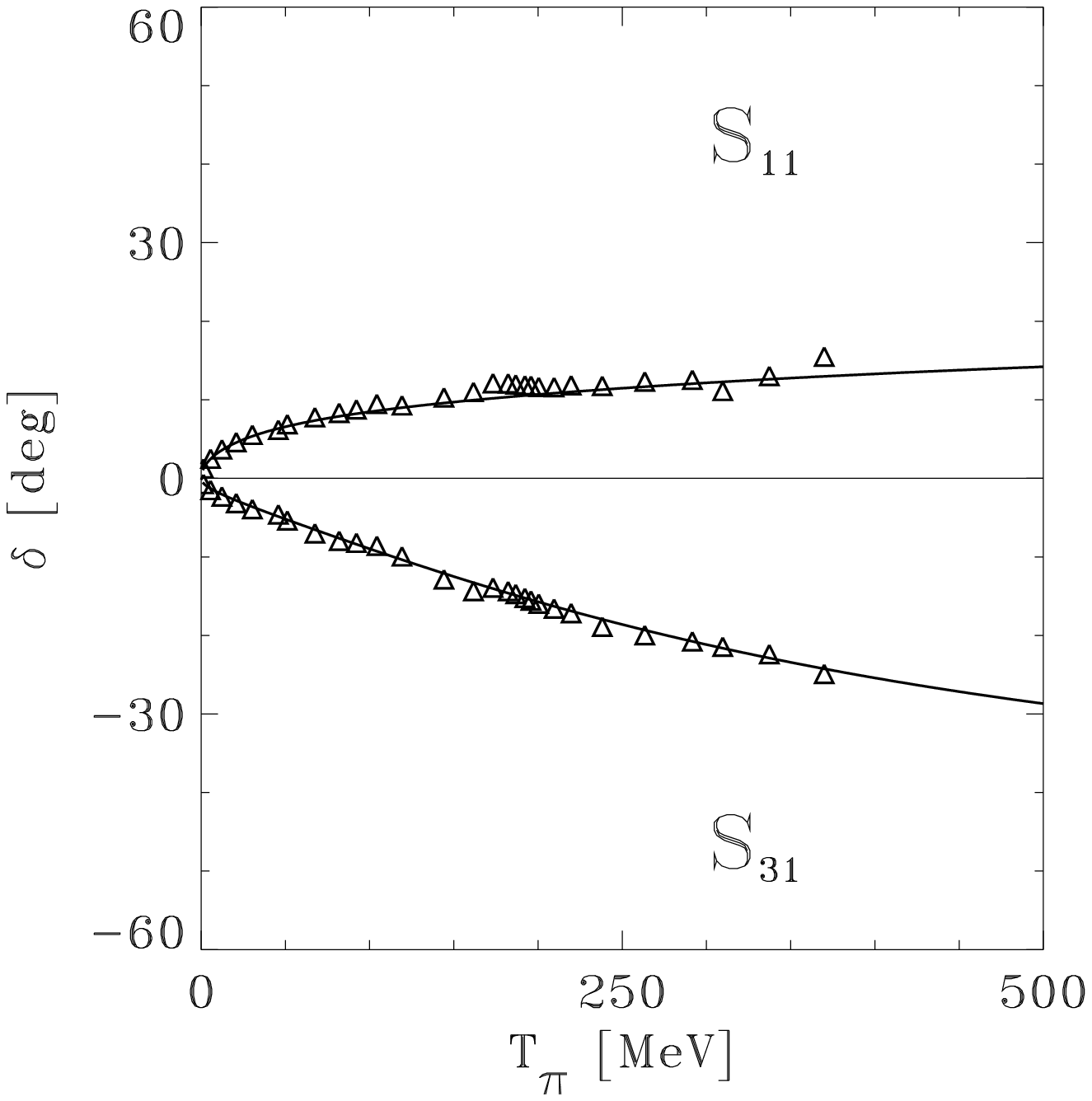,width=15.0cm}
\end{picture}
\caption{$\pi N$ SM95 \protect\cite{arnpin} (coinciding with the solid line) and
KH80  \protect\cite{koc80} (triangles) data and their 
reproduction  by the inversion potentials  (solid line) for the
$S_{11}$ and $S_{31}$ channels.}
\end{figure}

\begin{figure}\centering
\begin{picture}(15.0,15.0)(0.0,0.0)
\epsfig{figure=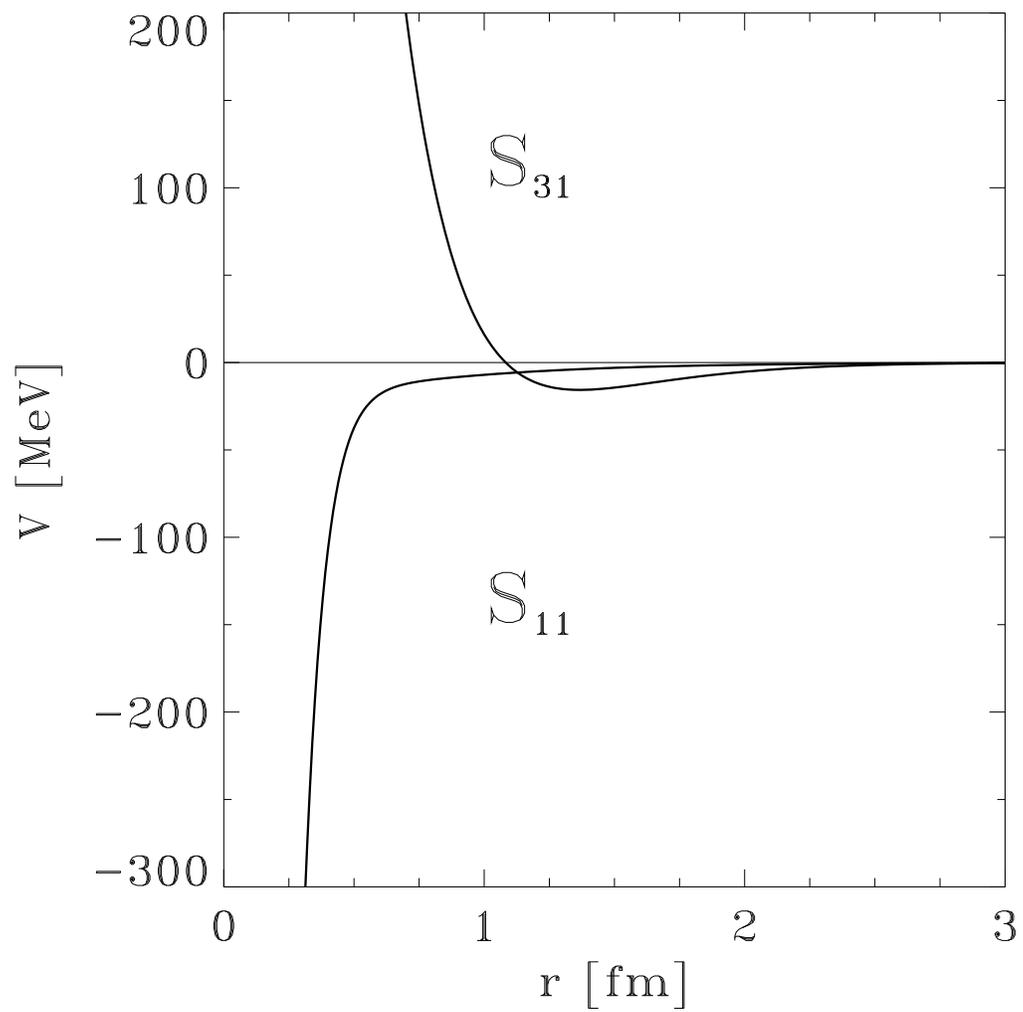,width=15.0cm}
\end{picture}
\caption{$S_{11}$ and $S_{31}$ inversion potentials.}
\end{figure}

\begin{figure}\centering
\begin{picture}(15.0,15.0)(0.0,0.0)
\epsfig{figure=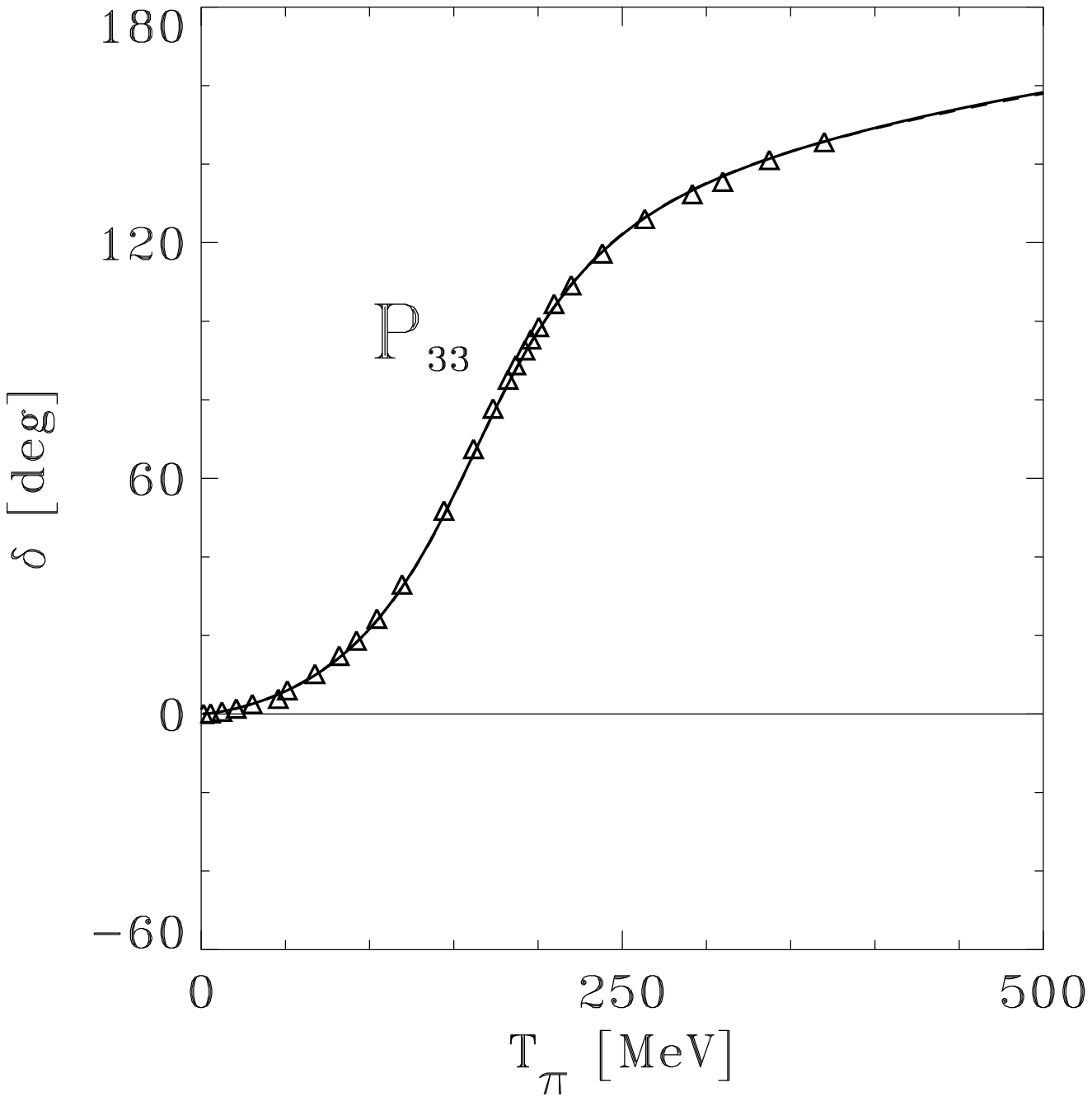,width=15.0cm}
\end{picture}
\caption{$\pi N$ SM95 \protect\cite{arnpin} (coinciding with the solid line) and
KH80  \protect\cite{koc80} (triangles) data and their 
reproduction  by the inversion potential  (solid line) for the
$P_{33}$ channel.}
\end{figure}

\begin{figure}\centering
\begin{picture}(15.0,15)(0.0,0.0)
\epsfig{figure=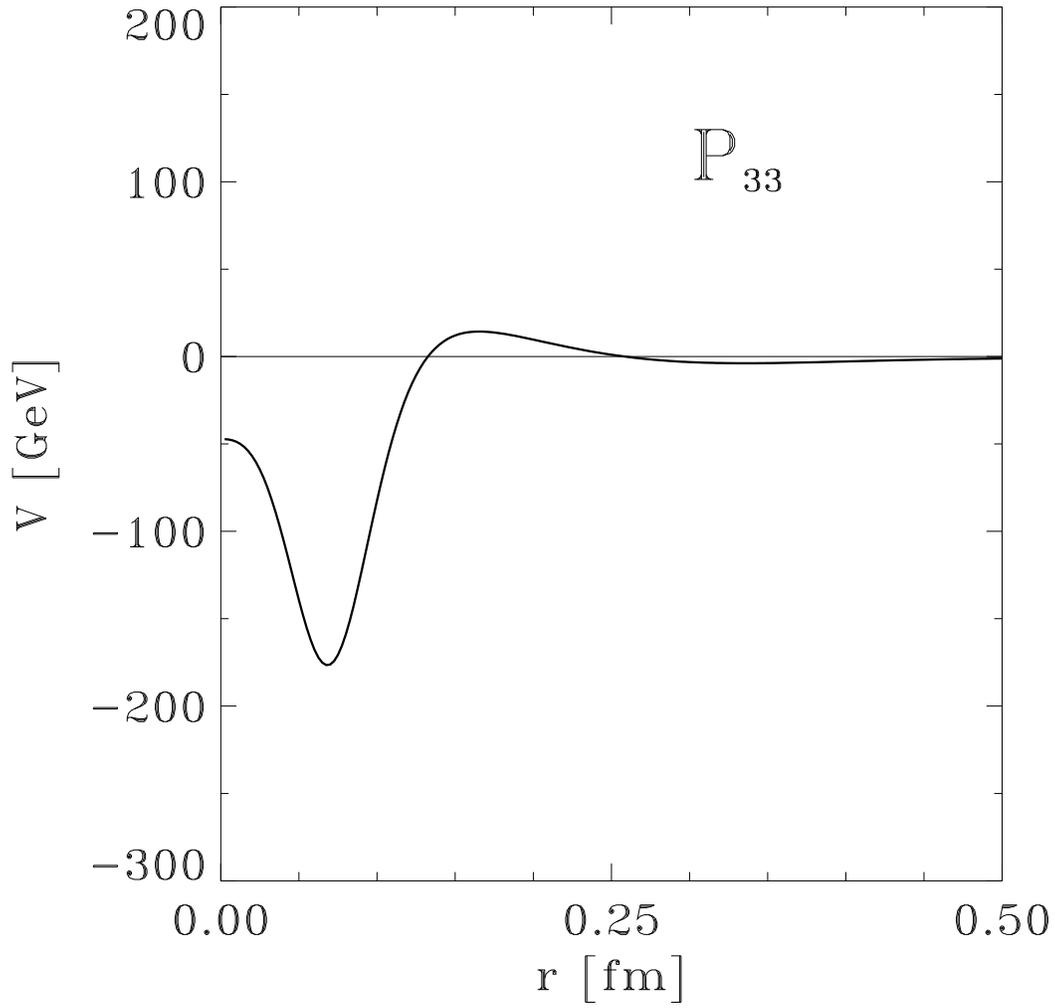,width=15.0cm}
\end{picture}
\caption{$P_{33}$ channel inversion potential.}
\end{figure}

\begin{figure}\centering
\begin{picture}(15.0,15.0)(0.0,0.0)
\epsfig{figure=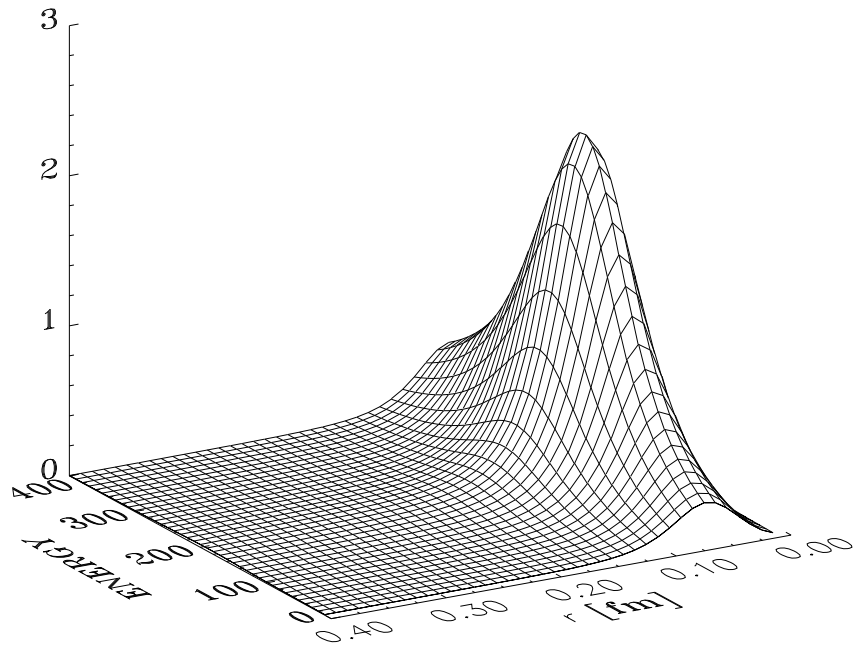,width=15.0cm}
\end{picture}
\caption{$\pi N$ $P_{33}$ channel radial probability distribution as function
of energy, $T_{lab}$.}
\end{figure}

\begin{figure}\centering
\begin{picture}(15.0,15.0)(0.0,0.0)
\epsfig{figure=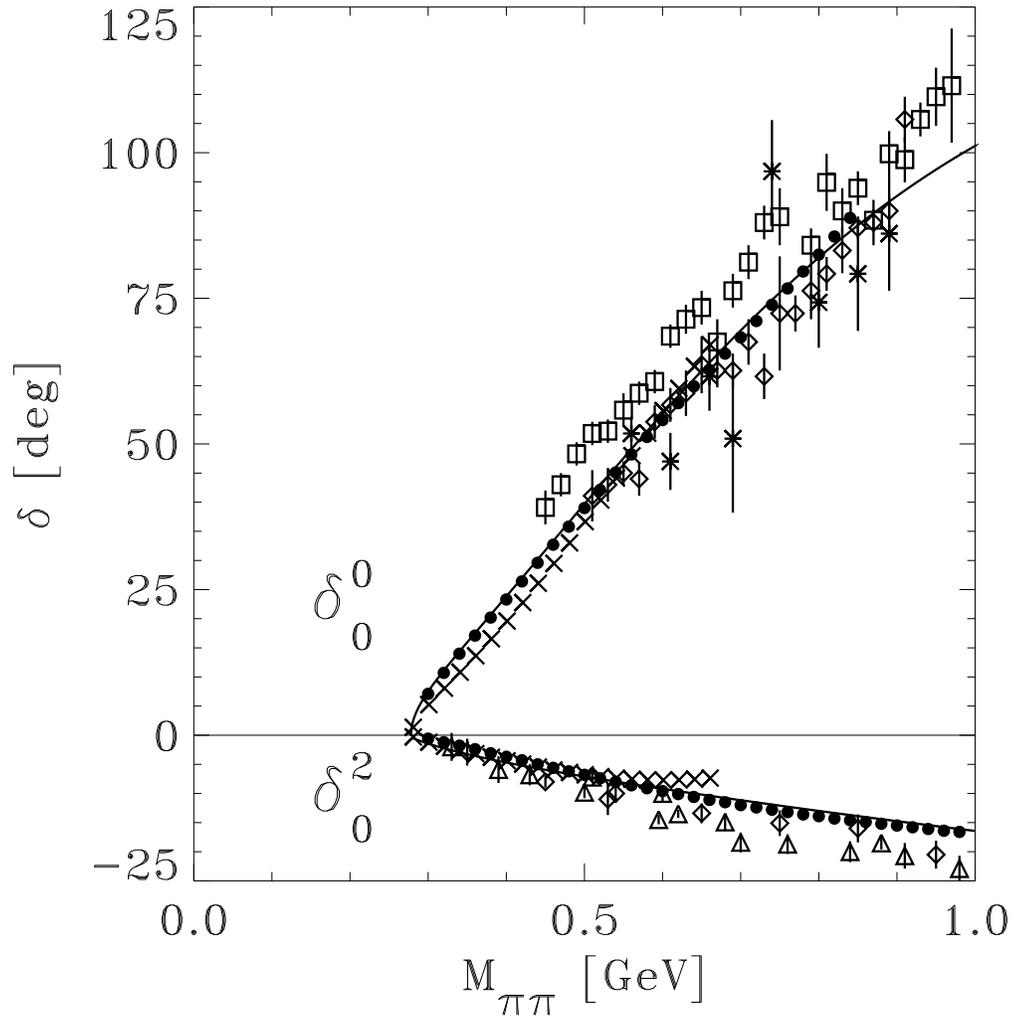,width=15.0cm}
\end{picture}
\caption{$\pi\pi$ $L=0$, $T=0$ and 2 phase shifts. 
Froggatt data \protect\cite{fro77}
and interpolation (dots and solid line), $\chi$PT \protect\cite{gas83}
(crosses), Estabrooks et al. \protect\cite{est74} (boxes),
Grayer et al. \protect\cite{gra74} (diamonds),
M\"anner \protect\cite{mae74} (triangles)
and Baillon et al. \protect\cite{bai72} (asterixes).}
\end{figure}

\begin{figure}\centering
\begin{picture}(15.0,15.0)(0.0,0.0)
\epsfig{figure=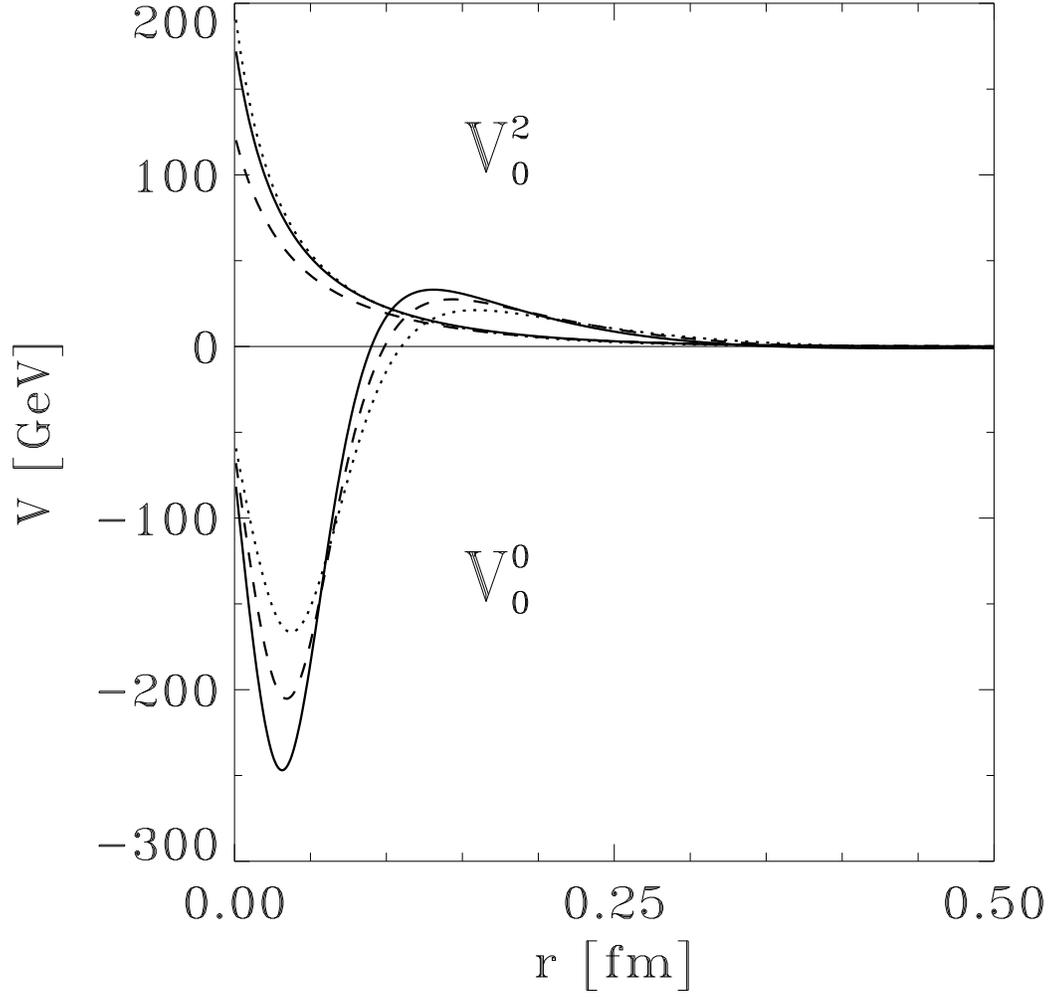,width=15.0cm}
\end{picture}
\caption%
{$\pi\pi$ $L = 0$, $T=0$ and 2 inversion potentials.
Froggatt (solid line), $\chi$PT (dashed)
and meson exchange \protect\cite{loh90} (dotted).}
\end{figure}

\begin{figure}\centering
\begin{picture}(15.0,15.0)(0.0,0.0)
\epsfig{figure=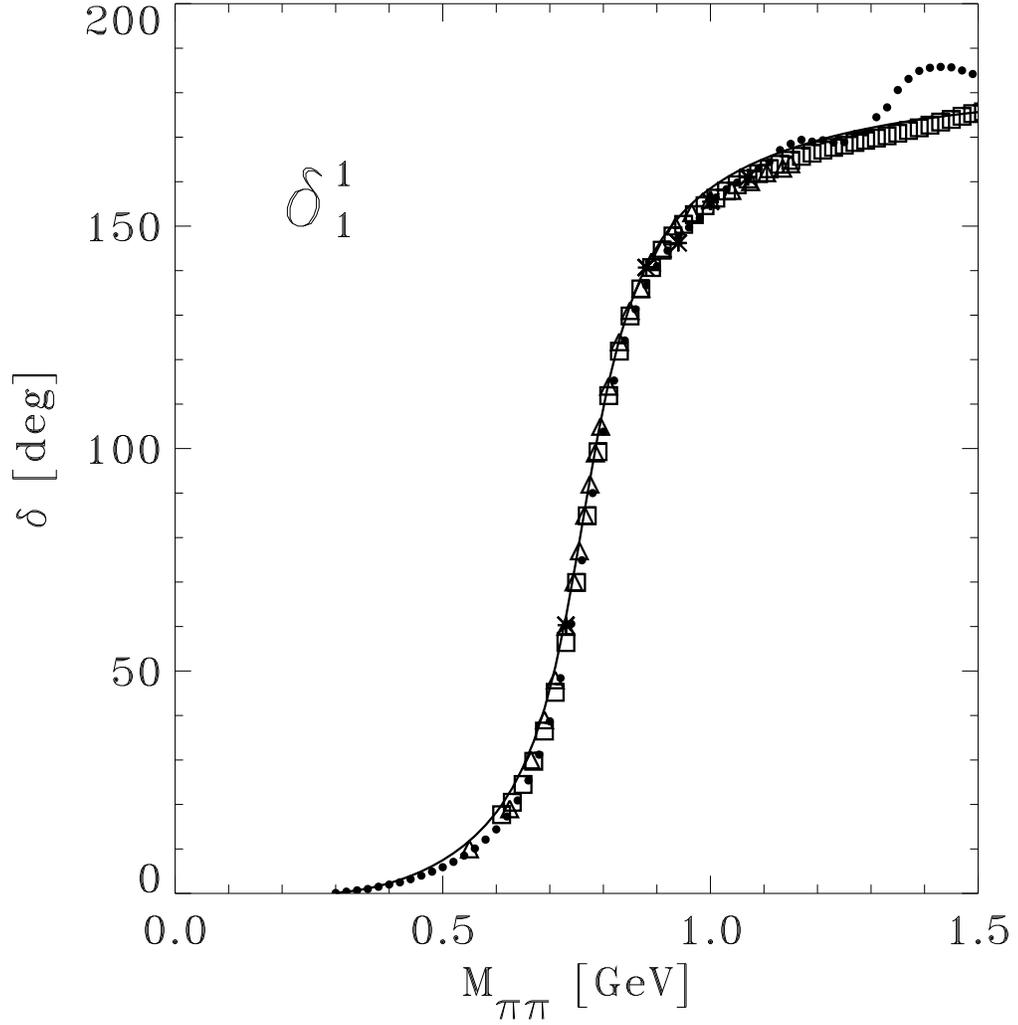,width=15.0cm}
\end{picture}
\caption{$\pi\pi$ $L = 1$, $T=1$ phase shifts.
Data from
Froggatt et al. \protect\cite{fro77} (dots),
 Ochs \protect\cite{och73} (squares), 
Protopopescu et al. \protect\cite{pro73} (triangles) and Deo et al.
\protect\cite{deo82} (asterixes). Inversion result (solid line).}
\end{figure}

\begin{figure}\centering
\begin{picture}(15.0,15.0)(0.0,0.0)
\epsfig{figure=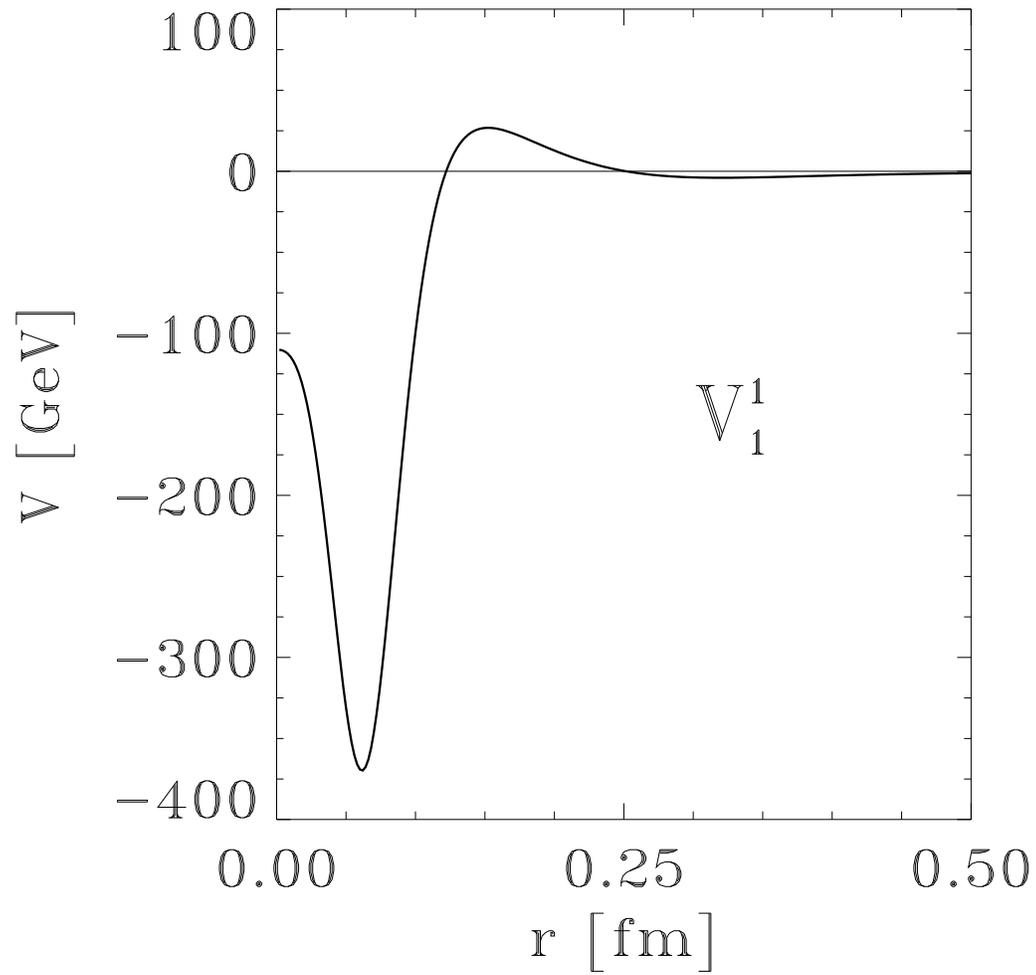,width=15.0cm}
\end{picture}
\caption{$\pi\pi$ $L = 1$, $T = 1$ inversion potential.}
\end{figure}

\begin{figure}\centering
\begin{picture}(15.0,15.0)(0.0,0.0)
\epsfig{figure=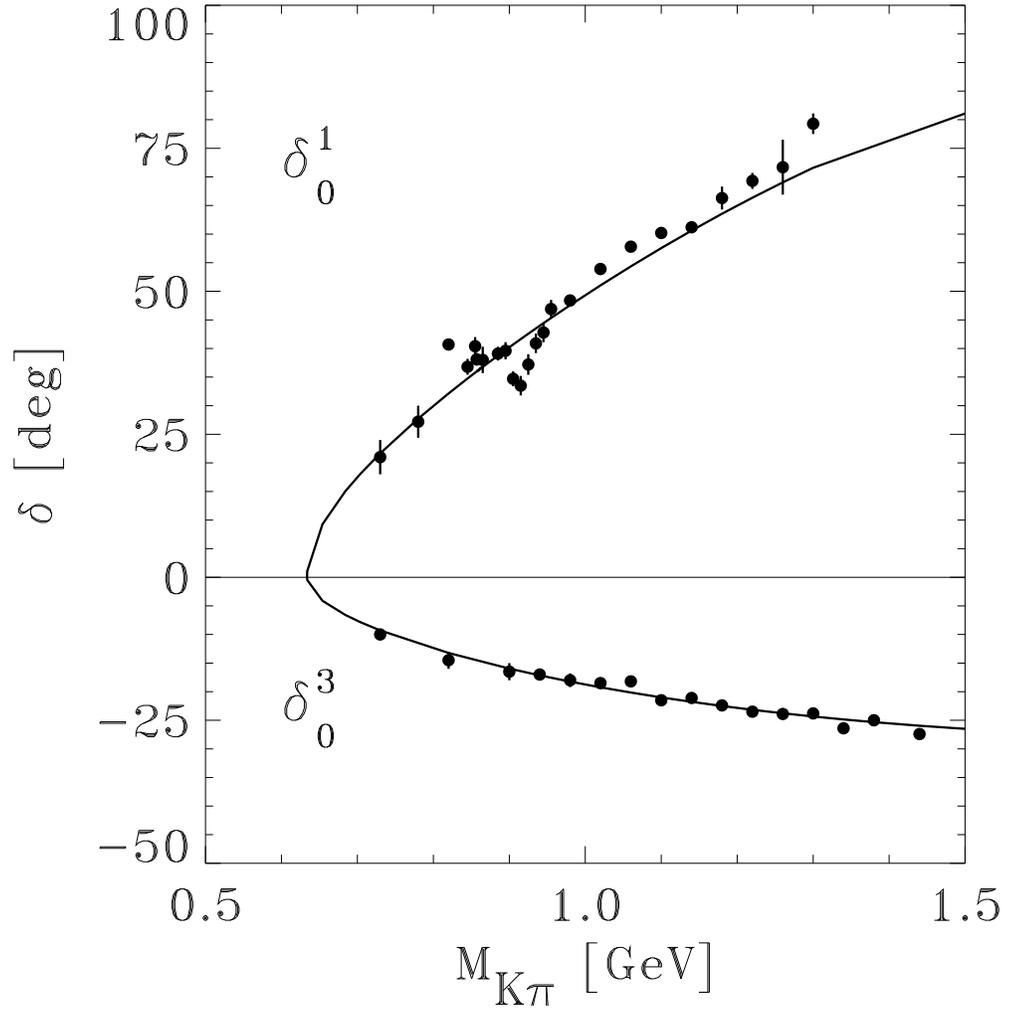,width=15.0cm}
\end{picture}
\caption{$K\pi$ $L = 0$, $T=1/2$ and $3/2$ phase shifts. 
Data from Estabrooks et al. \protect\cite{esta78}.}
\end{figure}

\begin{figure}\centering
\begin{picture}(15.0,15.0)(0.0,0.0)
\epsfig{figure=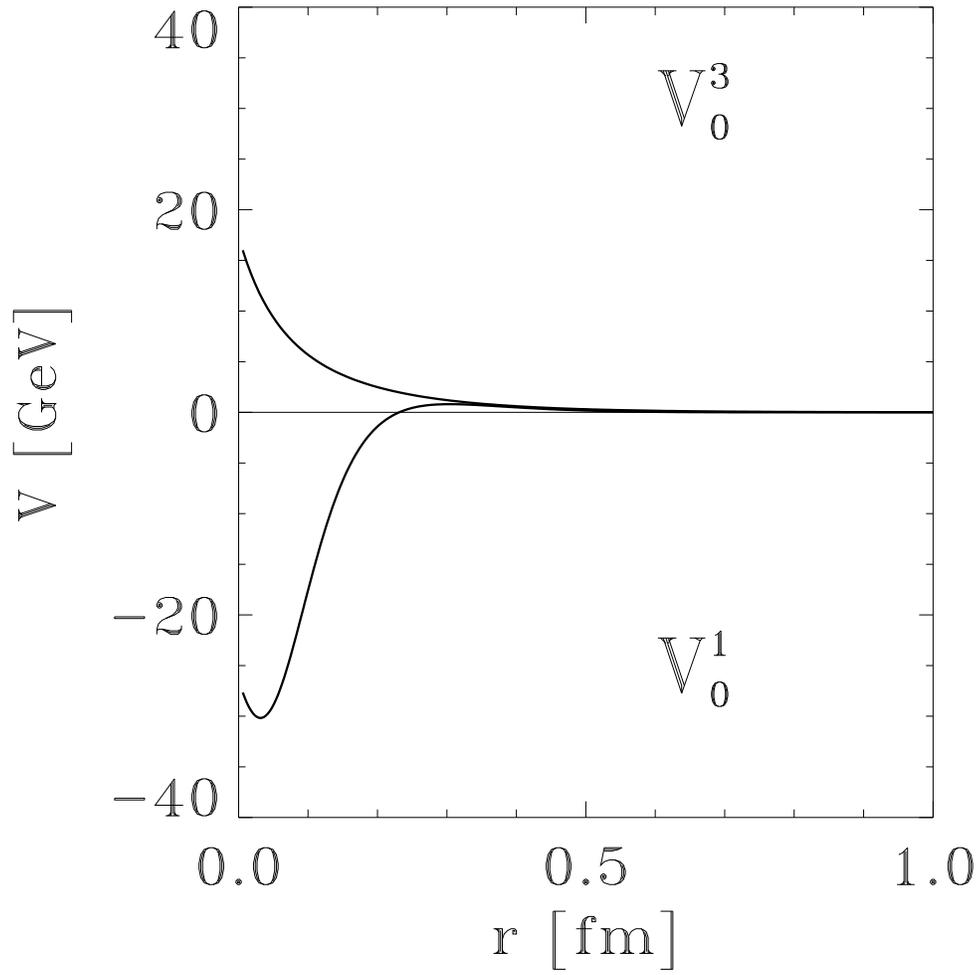,width=15.0cm}
\end{picture}
\caption%
{$K\pi$ $L = 0$, $T=1/2$ and $3/2$ inversion potentials.}
\end{figure}

\begin{figure}\centering
\begin{picture}(15.0,15.0)(0.0,0.0)
\epsfig{figure=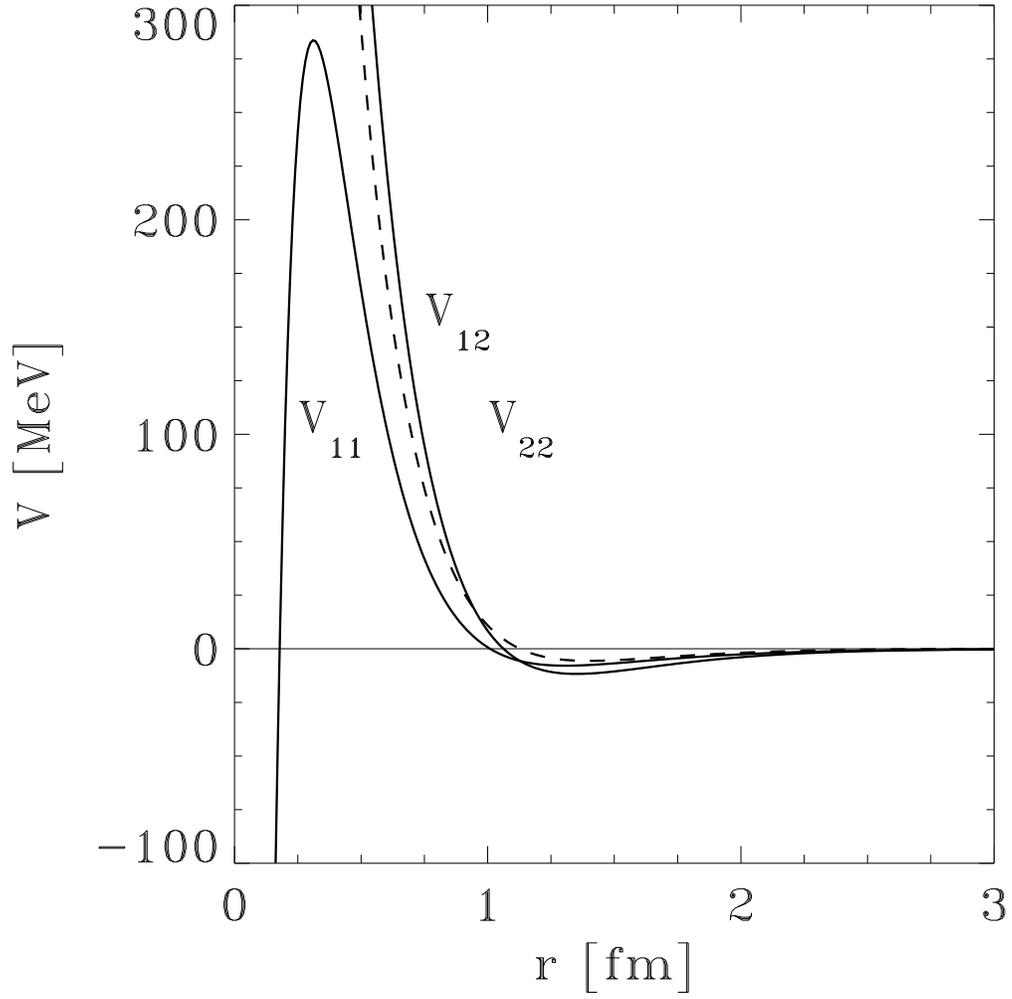,width=15.0cm}
\end{picture}
\caption{$\pi N$ hadronic potential matrix.}
\end{figure}

\begin{figure}\centering
\begin{picture}(15.0,15.0)(0.0,0.0)
\epsfig{figure=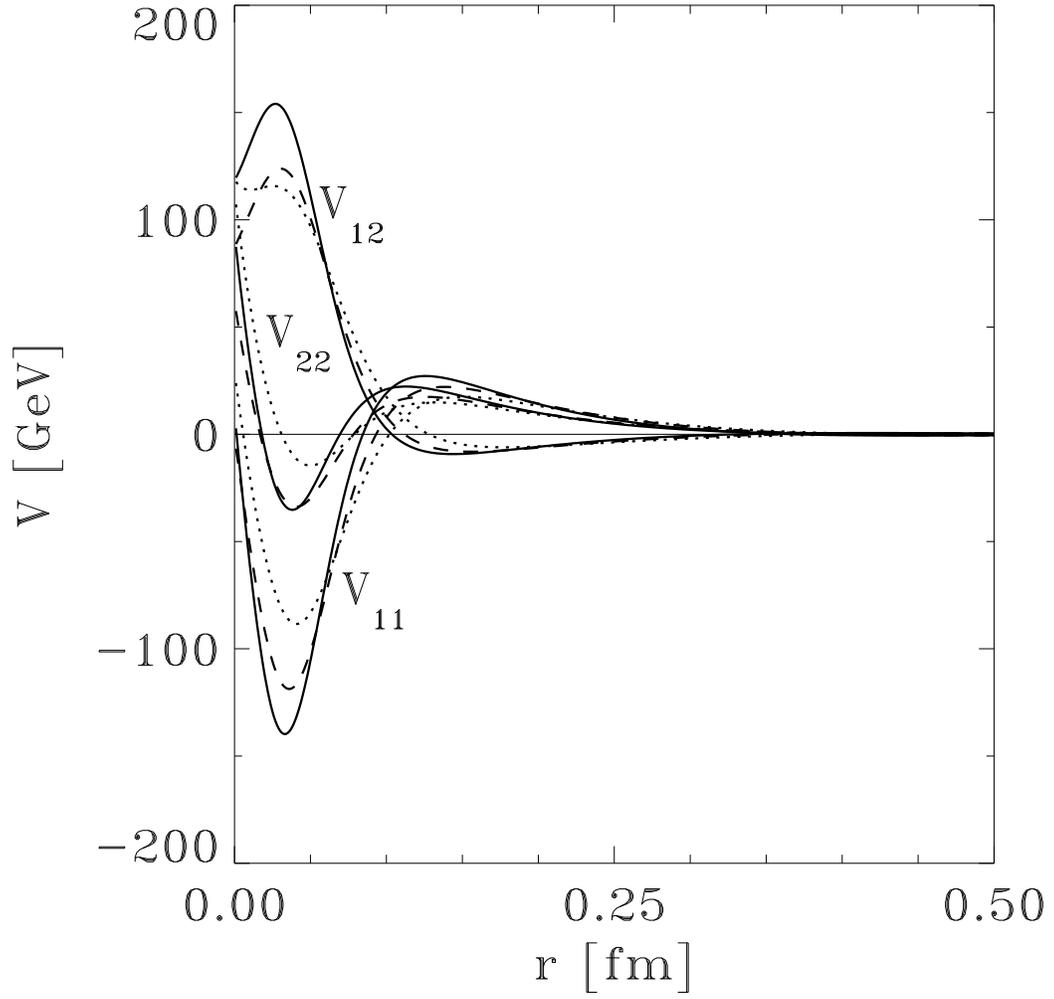,width=15.0cm}
\end{picture}
\caption%
{$\pi\pi$ hadronic potential matrix. Notation, see Fig. 7.}
\end{figure}

\begin{figure}\centering
\begin{picture}(15.0,10.0)(0.0,0.0)
\epsfig{figure=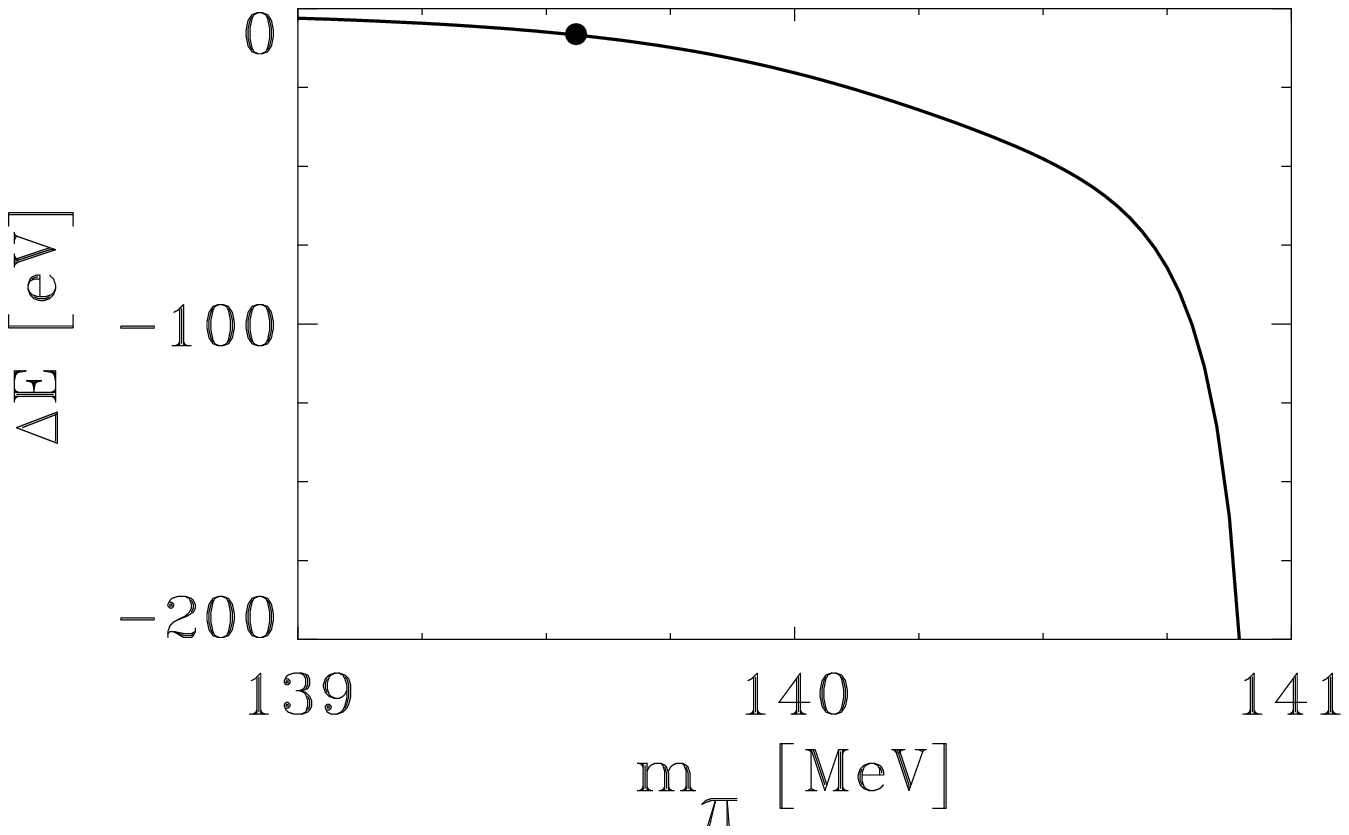,width=15.0cm}
\end{picture}
\begin{picture}(15.0,10.0)(0.0,0.0)
\epsfig{figure=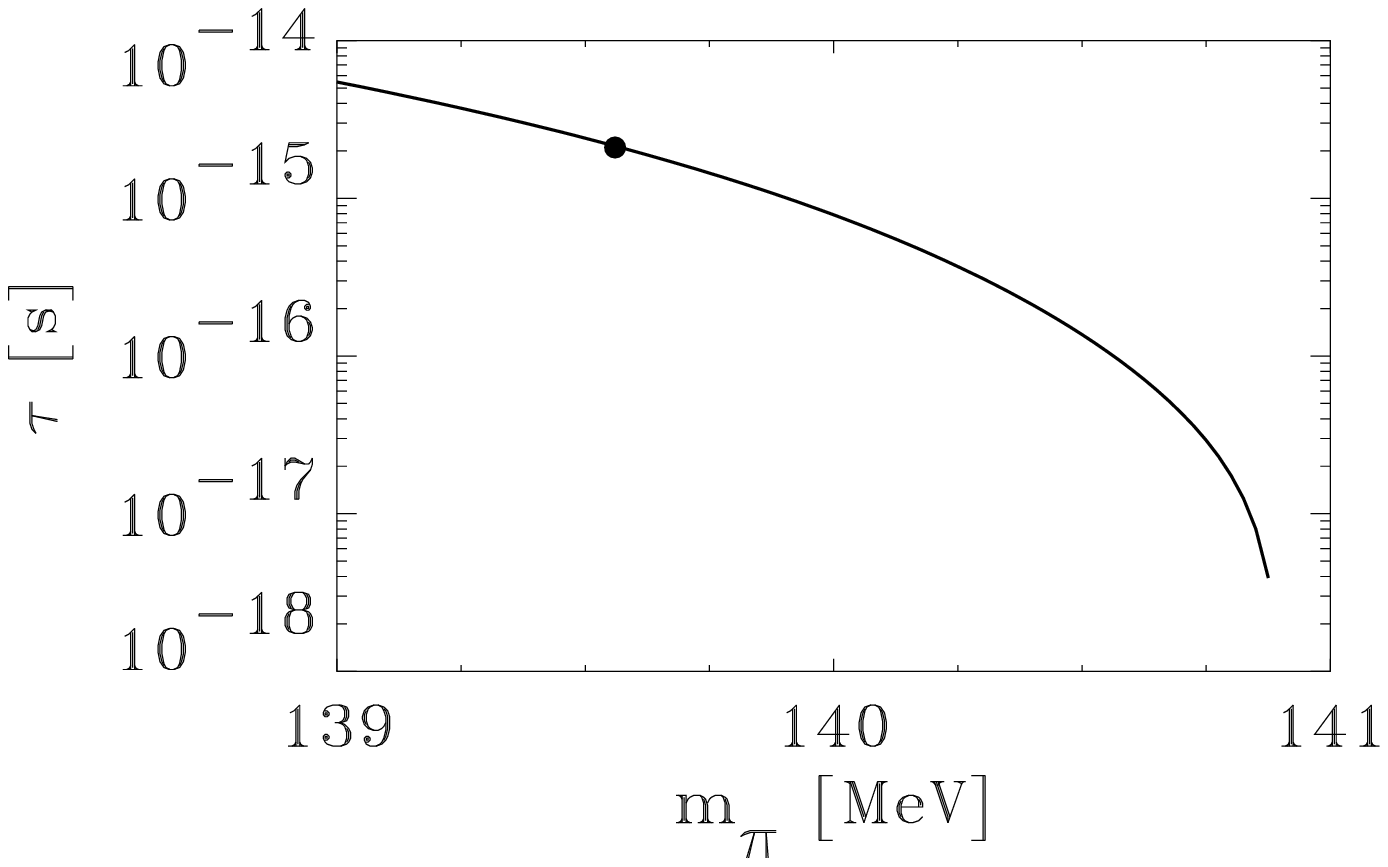,width=15.0cm}
\end{picture}
\caption{\PP\ shift (above) and width (below) as a function 
of the effective mass. The physical value $m_{\pi^+}=139.5676$ MeV is 
emphasized.}
\label{figure1}
\end{figure}

\begin{figure}\centering
\begin{picture}(15.0,15.0)(0.0,0.0)
\epsfig{figure=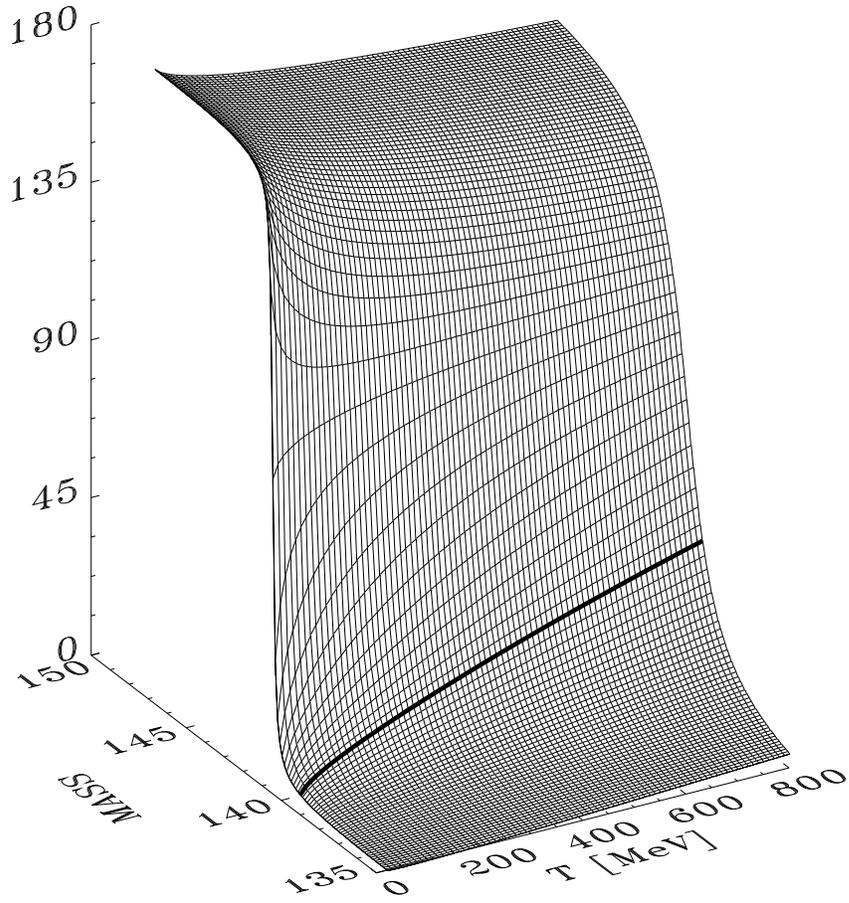,width=15.0cm}
\end{picture}
\caption{$\pi\pi$ $L=0$, $T=0$ phase shift as a function of energy ($T_{lab}$)
and effective mass. The physical value $m_{\pi^+}=139.5676$ MeV is 
emphasized.}
\end{figure}

\begin{figure}\centering
\begin{picture}(12.0,7.0)(0.0,0.0)
\epsfig{figure=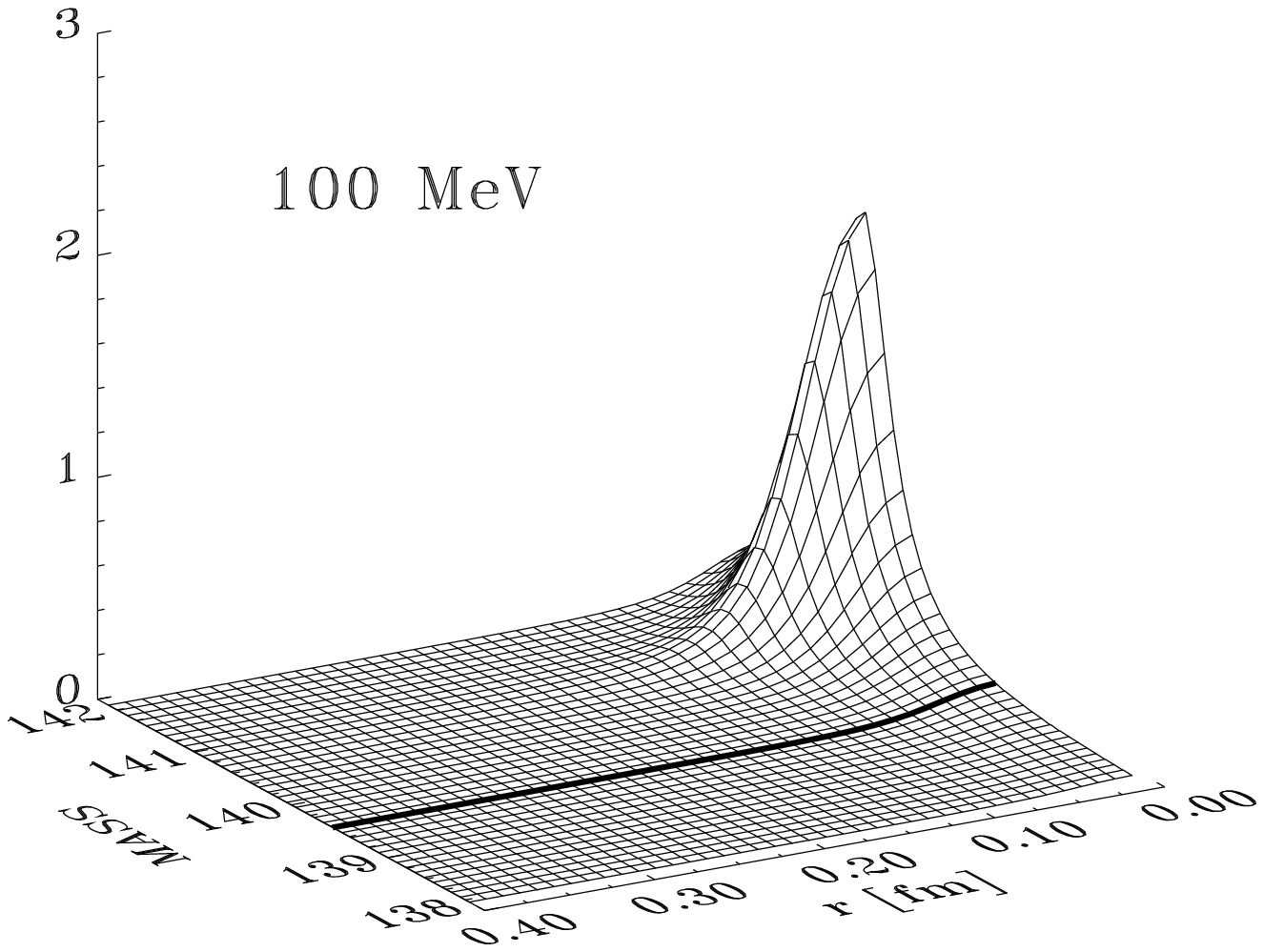,width=12.0cm}
\end{picture}
\\
\begin{picture}(12.0,7.0)(0.0,0.0)
\epsfig{figure=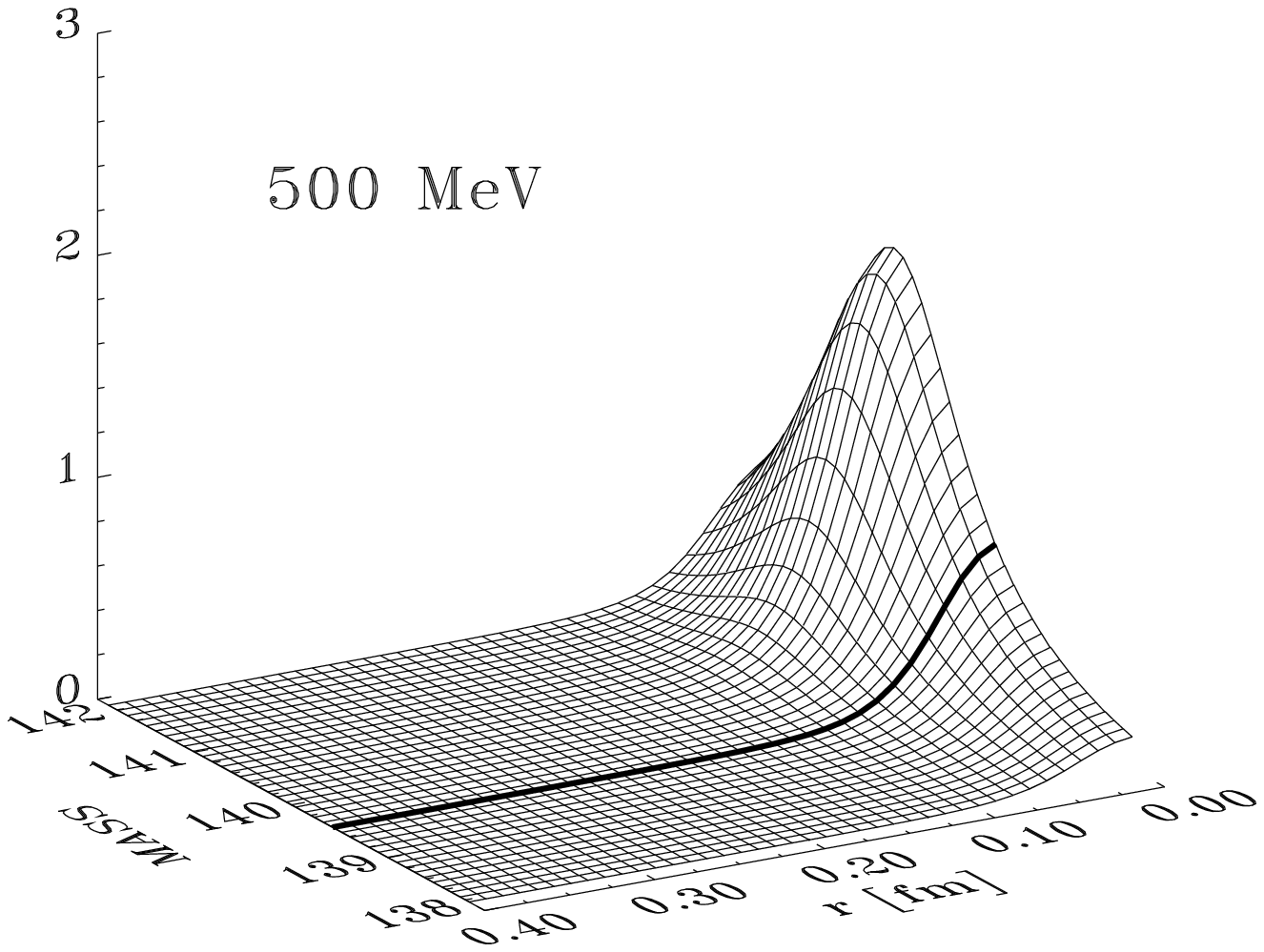,width=12.0cm}
\end{picture}
\\
\begin{picture}(12.0,7.0)(0.0,0.0)
\epsfig{figure=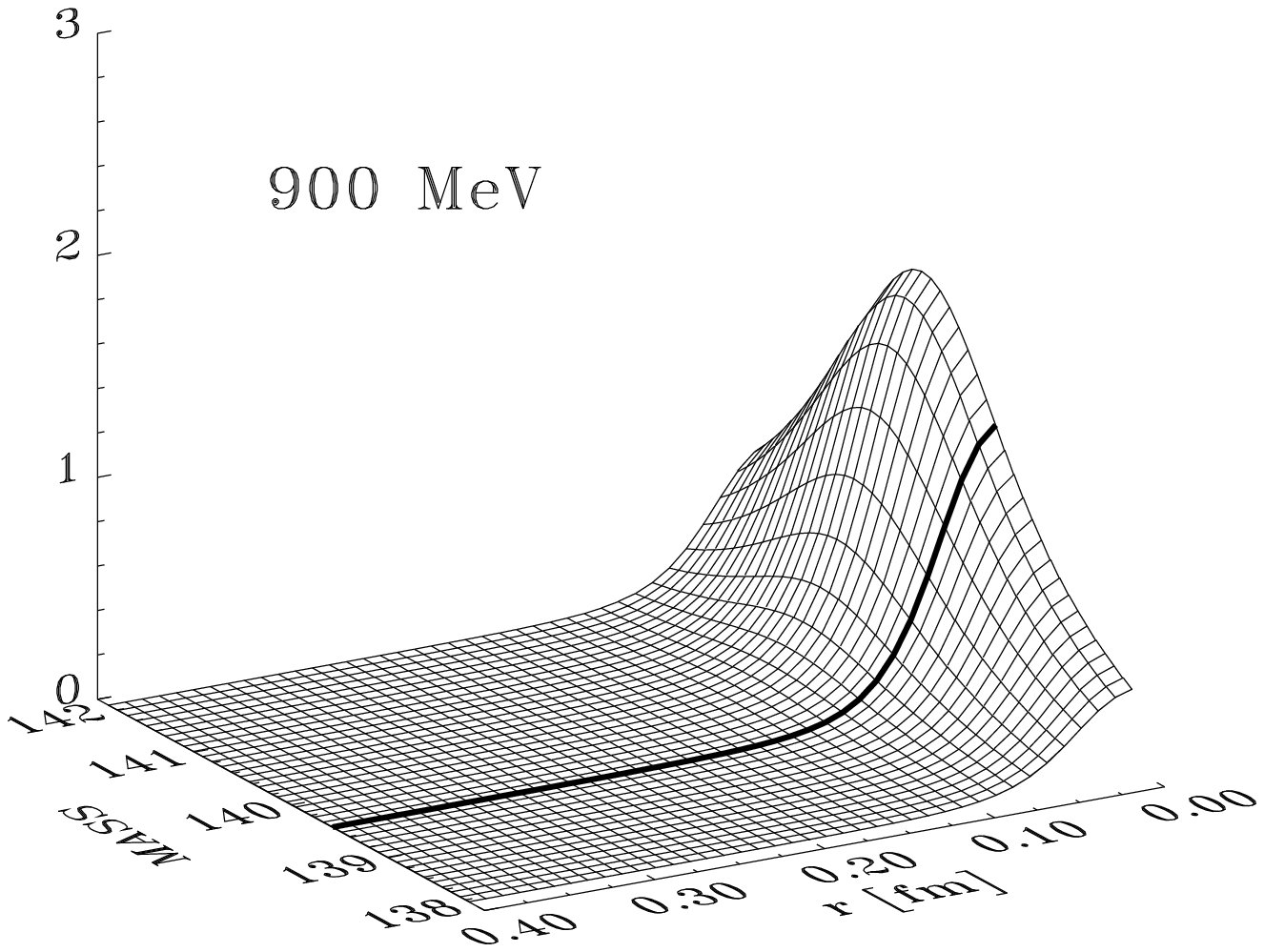,width=12.0cm}
\end{picture}
\caption{$\pi \pi$ $L=0$, $T=0$ radial probability distribution as a function
of effective mass given for three energies ($T_{lab} =$ 100, 500 and 900 MeV).
The physical value $m_{\pi^+}=139.5676$ MeV is emphasized.}
\end{figure}

\begin{figure}\centering
\begin{picture}(15.0,15.0)(0.0,0.0)
\epsfig{figure=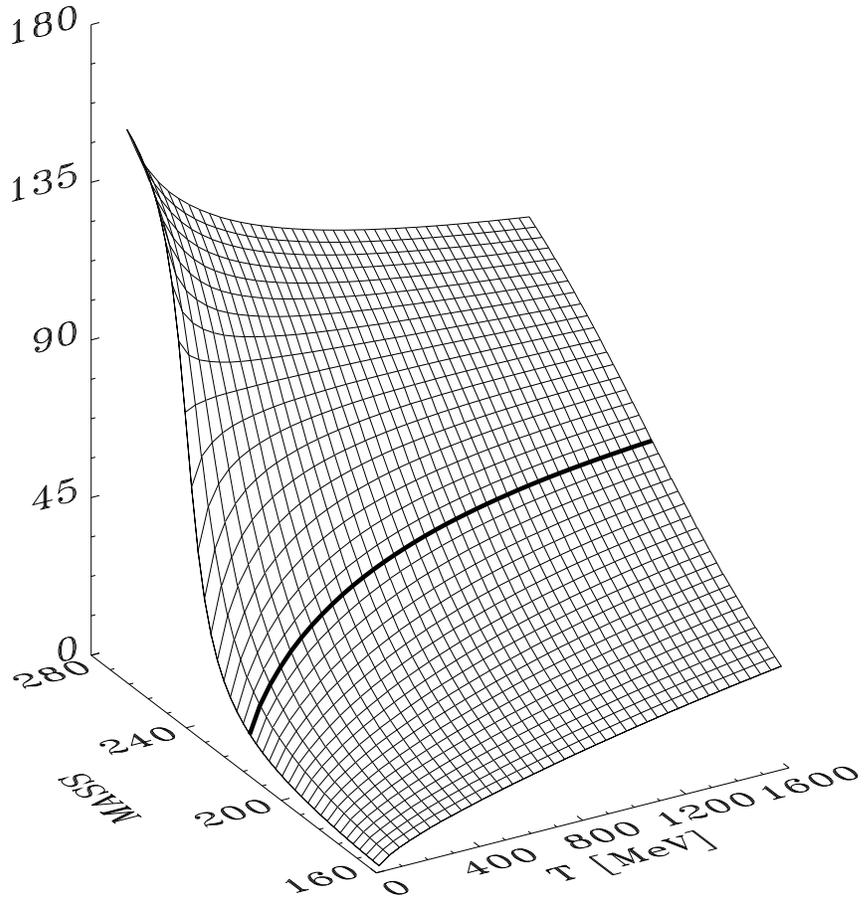,width=15.0cm}
\end{picture}
\caption{$K\pi$ $L=0$, $T=1/2$ phase shift as a function of energy ($T_{lab}$)
and effective mass. The physical value $2\mu_{K\pi}=215.94$ MeV is
emphasized.}
\end{figure}


\end{document}